\documentclass[12pt,a4paper]{article}

\usepackage{amssymb}
\usepackage{epsf}
\usepackage{amsmath,amsopn}

\newcommand{\QATOP}[2]{\genfrac{}{}{0pt}{}{#1}{#2}}

\begin{document}

\title{\textbf{Hydrodynamics of phase transition fronts and the speed of sound in the plasma}}
\author{{\large Leonardo Leitao\thanks{%
Fellow of CONICET, Argentina.
E-mail address: lleitao@mdp.edu.ar}~ and Ariel M\'{e}gevand\thanks{%
Member of CONICET, Argentina. E-mail address: megevand@mdp.edu.ar}} \\
{\normalsize \textit{IFIMAR (CONICET-UNMdP)}}\\
{\normalsize \textit{Departamento de F\'{\i}sica, Facultad de Ciencias
Exactas y Naturales, }}\\
{\normalsize \textit{UNMdP, De\'{a}n Funes 3350, (7600) Mar del Plata,
Argentina }}}
\date{}
\maketitle

\begin{abstract}
The growth of bubbles in cosmological first-order phase transitions involves
nontrivial hydrodynamics. For that reason, the study of the propagation of
phase transition fronts often requires several approximations. A frequently
used approximation consists in describing the two phases as being composed only of radiation and vacuum energy (the so-called bag equation of state).
We show that, in realistic models, the speed of sound in  the low-temperature phase is generally smaller than that of radiation, and we study the hydrodynamics in such a situation.
We find in particular that a new kind of hydrodynamical solution may be possible, which does not arise in the bag model. We obtain analytic results for the efficiency of the transfer of latent heat to bulk motions of the plasma, as a function of the speed of sound in each phase.
\end{abstract}

\section{Introduction \label{intro}}

First-order phase transitions of the Universe generically lead to interesting
phenomena and may have several observable consequences, such as topological
defects \cite{vs94}, magnetic fields \cite{gr01}, the baryon asymmetry of
the Universe \cite{ckn93}, inhomogeneities \cite{w84,ma05}, or gravitational
waves (see, e.g., \cite{kt93,kkt94,gw}). Most of these cosmological remnants depend on the disturbance produced  in the  plasma
by the nucleation and expansion of bubbles.
This fact has led to extensive studies of the hydrodynamics associated with
the motion of bubble walls (see, e.g., \cite{s82,gkkm84,mp89,ms09}), as
well as investigations of the microphysics responsible for the friction of these walls with the plasma \cite{fric}.

Studying the hydrodynamics of these phase transition fronts is a
difficult task, and several approximations are generally needed in order to
simplify the analysis. Thus, it is usual to study the propagation of a single phase transition front, disregarding some global aspects of the dynamics of the phase transition (e.g., the presence of other bubbles or the cooling of the Universe).
In addition, the wall is assumed to move at a constant velocity. This is
in general a good approximation since a terminal velocity is  reached in a very short time after nucleation.
Other simplifications include
considering symmetric bubble walls, e.g., spherical or planar walls. The hydrodynamics of planar walls is easier to treat than that of spherical walls. The results are qualitatively similar but quantitatively different \cite{lm11}. In the first stages of their expansion, the bubbles are certainly spherical. However, once bubbles collide this symmetry is lost. To treat the case of colliding bubbles, the envelop approximation \cite{kt93} is often used in the calculation of gravitational waves. This approximation assumes spherical (overlapping) bubbles, neglects the  regions in which two or more bubbles overlap, and follows only the parts of walls which have not collided yet. However, when bubbles meet and coalesce, their walls straighten due to surface tension \cite{w84}. Thus, for a wall which envelops a system of several bubbles, the planar wall approximation may be as good as the spherical approximation.
Assuming an infinitely thin wall is also a good approximation for the treatment of hydrodynamics. The phase on each side of this interface can be described using a phenomenological equation of state (EOS).
A frequently used approximation for the equation of state is given by the
bag EOS, which simplifies considerably the calculations and sometimes (e.g., in the case of a planar wall) even leads to analytical results.

The bag EOS corresponds to having only radiation and vacuum energy in both
the high- and low-temperature phases, which we shall denote with
a $+$ and a $-$, respectively.
Therefore, this EOS depends on a few free parameters, namely, the vacuum energy densities $\epsilon_\pm$ (in general, $\epsilon
_{-}=0$ is assumed) and the radiation constants $a_\pm$, which are proportional to the Stefan-Boltzmann constant and the number of degrees of freedom in each phase.
Besides being useful for analyzing general properties of the phase transition, the results obtained using the bag EOS can be applied to realistic models by calculating the values of the constants
$\epsilon _{\pm }$ and $a_{\pm }$ for such models. To do that, one has to
identify, for a given model, the vacuum energy density
and the radiation constant in each phase.
Since in general the
model will not consist of just vacuum energy and radiation, it will be
necessary to define effective constants $\epsilon_\pm,a_\pm$ which will in fact depend on the temperature $T$.
Needless to say, the
definitions of $a_{\pm }(T)$ and $\epsilon _{\pm }(T)$
are ambiguous for a general system.

Nevertheless, in general one may set the values of the bag parameters so
as to give the desired values of some relevant quantities (e.g., the latent
heat, the critical temperature, etc.).
However, the few free parameters of the bag EOS may fall short of
describing all the desired features of the model under study. In
particular, the hydrodynamics associated to the motion of bubble walls
depends on the speed of sound in the plasma, which is given by
\begin{equation}
c_s^{2}\equiv \frac{\partial p}{\partial e}  \label{csdef}
\end{equation}
(we use natural units with $c=\hbar=k_B=1$).
For the bag model, the speed of sound is that of radiation in both phases, $c_{\pm}=1/\sqrt{3}$. For a realistic model, however, $c_s$ will be different in each phase. Moreover, in general we will have a temperature-dependent functions $c_\pm(T)$. As a consequence, the  sound velocity will vary in space and time during the phase transition.

In this paper, we shall discuss to what extent the speed of sound may depart from the radiation value $c_s=1/\sqrt{3}$ in a realistic model. As we shall see, in the low-temperature phase the speed of sound $c_-$ may be significantly smaller than that value. We shall also investigate the implications of having a different speed of sound in each phase.
For that aim, we shall consider a family of models
which can be regarded as the simplest generalization of the bag EOS. We obtain this phenomenological EOS by requiring it to give a constant speed of sound. The model is still simple enough to obtain analytical results for planar walls. We shall calculate, in particular, the kinetic energy in bulk motions of the fluid, after finding the different kinds of hydrodynamic
solutions for the propagation of a phase transition front.

The plan is the following. In the next section we study the possible
values of the speed of sound in a physical model and we discuss the
shortcomings of using phenomenological equations of state as an approximation. In Sec. \ref{hidrogral} we review the hydrodynamics involved in the propagation of phase transition fronts and we discuss on some ambiguities and some misleading definitions in the literature.
In Sec. \ref{model} we introduce our model family.
We analyze the different kinds of phase transitions described by this model, and we study the hydrodynamics for each case.
For the case $c_-<c_+$, we find a solution which
does not arise for the bag EOS, namely, a Jouguet detonation which is
subsonic with respect to the fluid in front of it, and as a consequence is
preceded by a shock wave.
We also calculate the fraction
of the energy released at the phase transition which goes into bulk motions
of the fluid. In Sec. \ref{effic} we study the dependence
of this quantity on the speed of sound. We summarize
our conclusions in Sec. \ref{conclu}, and we provide analytical results in two appendices: in App. \ref{flprof} we give the equations for the fluid profiles, and in App. \ref{intrar} we provide the integral of the kinetic energy density in the rarefaction region.

\section{The equation of state \label{EOSs}}

The free energy density $\mathcal{F}$ of a system may depend on an order
parameter $\phi $. This generally happens in models with scalar fields,
where $\mathcal{F}$ is given by the finite-temperature effective potential,
and $\phi $ is given by the expectation value of one or more
of the scalar fields. The equilibrium state of the system corresponds
to a minimum of the free energy. As a consequence,
if the function $\mathcal{F}(\phi ,T)$ has more than one minimum  a cosmological phase transition may occur. A
first-order phase transition occurs when two such minima coexist, separated by a free energy barrier, in a
certain range of temperatures. One of
these minima is the absolute minimum at higher temperatures, while the other
one becomes the absolute minimum at lower temperatures. Let us denote these
minima $\phi _{+}\left( T\right) $ and $\phi _{-}\left( T\right) $,
respectively (we shall use in general a ``$+$'' index for the high-temperature
phase and a ``$-$'' index for the low-temperature phase). The critical
temperature $T_{c}$ is defined as that at which the two minima have exactly
the same free energy density. Therefore, for $T>T_{c}$ the system is in the
stable phase given by the minimum $\phi _{+}$, while for $T<T_{c}$ this
phase becomes metastable. The phase transition typically develops via the
nucleation and expansion of bubbles at a given temperature $T_{n}$ which is lower than $T_{c}$,
i.e., a certain amount of supercooling occurs before the phase transition
begins (see, e.g., \cite{nucl}).

The equation of state  can be derived from the equilibrium free energy
densities in each phase, given by $\mathcal{F}_{+}\left( T\right) \equiv \mathcal{F}
\left( \phi _{+}\left( T\right) ,T\right) $ and $\mathcal{F}_{-}\left(
T\right) \equiv \mathcal{F}\left( \phi _{-}\left( T\right) ,T\right) $.
Thus, the pressure is given by $p=-\mathcal{F}$, the entropy density by $
s=dp/dT$, the energy density by $e=Ts-p$, and the enthalpy by $w=e+p=Ts$. At
the critical temperature the two phases have the same pressure, $p_{+}\left(
T_{c}\right) =p_{-}\left( T_{c}\right) $. On the other hand, the energy,
entropy and enthalpy are different in each phase, and we will have
discontinuities at the phase transition. The latent heat is defined as the
energy density discontinuity at the critical temperature, $L=\Delta e\left(
T_{c}\right) =\Delta w\left( T_{c}\right) =T_{c}\Delta s\left( T_{c}\right) $.
For a bubble expanding at $T=T_{n}$, an energy density $\Delta
e(T_{n})\simeq L$ is released at the phase transition fronts (bubble walls).

A simple approximation for the equation of state in each phase is given by the bag EOS,
\begin{equation}
\begin{array}{ccc}
e_{+}=a_{+}T^{4}+\epsilon _{+}, &  & e_{-}=a_{-}T^{4}+\epsilon _{-},
\\
p_{+}=\frac{1}{3}a_{+}T^{4}-\epsilon _{+} &  & p_{-}=\frac{1}{3}
a_{-}T^{4}-\epsilon _{-},
\end{array}
\label{EOSbag}
\end{equation}
which can be derived from the free energy density
\begin{equation}
\mathcal{F}_\pm(T)=\epsilon _{\pm}-\frac{1}{3}a_{\pm}T^{4}. \label{Fbag}
\end{equation}
This equation of state is based on the bag model for hadrons \cite{bagmodel}. In that case, Eq. (\ref{EOSbag}) describes a first-order QCD phase transition\footnote{Although the quark-hadron transition was initially assumed to be a first-order phase transition, lattice calculations \cite{orderQCD} showed that this transition is in fact a crossover.}.
Physically, the approximation
represented by Eq. (\ref{EOSbag})  corresponds to assuming
that the two phases  consist
of a gas of massless particles, each one with different numbers (and kinds) of particle species (namely, quarks and gluons in the $+$ phase, and pions in the $-$ phase). These numbers of degrees of freedom (d.o.f.) are proportional to the constants $a_\pm$. In this context, the constant $\epsilon_+$ is given by the bag constant $B$, and $\epsilon_-$ is assumed to vanish.

With the aim of simplifying the treatment of hydrodynamics, the bag EOS is often considered as an approximation to describe general phase transitions, including the electroweak phase transition (see, for instance, \cite{kkt94,s82,ekns10,kn11,ms12}). In the context of a Higgs mechanism, this EOS can be interpreted as follows. For $T>T_c$, the system is in a false vacuum and we have a certain number of massless d.o.f. Therefore, we have a vacuum energy density $\epsilon_+$ and a radiation energy density $a_+ T^4$. At $T=T_{c}$, a number of degrees of freedom (proportional to $\Delta a=a_+-a_-$) suddenly become very massive and disappear from the plasma. At the same time, a false vacuum energy density $
\Delta \epsilon =\epsilon_+-\epsilon_-$ is liberated.
In order to compare this approximation
with a realistic case, let us consider a simple system in which some of the
particles masses depend on a Higgs field $\phi$, and the free energy
density is given by the one-loop finite-temperature effective potential
\begin{equation}
\mathcal{F}(\phi ,T)=V(\phi )+V_{T}(\phi ),  \label{ftot}
\end{equation}
where $V(\phi )$ is the renormalized zero-temperature effective potential and $V_{T}(\phi )$ is the finite-temperature correction \cite{quiros}
\begin{equation}
V_{T}(\phi )= \sum_{i}(\pm g_{i})
T^{4}\int \frac{x^{2}dx}{2\pi ^{2}}\log \left[ 1\mp e^{-\sqrt{x^{2}+\left(
\frac{m_{i}}{T}\right) ^{2}}}\right] .  \label{f1loop2}
\end{equation}
where $g_{i}$ is the number of degrees of freedom of particle species $i$,
the upper sign corresponds to bosons and the lower sign to fermions, and $m_{i}$ is the mass of the particle.

For $m_{i}=0$, the species contributes a term
$- c_{i}g_{i}(\pi ^{2}/90)
T^{4}$ to $V_{T}(\phi )$,
where $c_{i}=1$ for bosons and $7/8$ for fermions. On the other hand, for $
m_{i}/T\gg 1$, the integral in Eq. (\ref{f1loop2}) is exponentially suppressed, and the species gives a vanishing contribution.
Thus,
the bag EOS is obtained in the limit in which all the particles are very
light in the high-temperature phase, i.e., $m_{i}(\phi _{+})/T\simeq 0$, while
some of the species acquire very large masses in the low-temperature phase, i.e., $
m_{i}(\phi _{-})/T\gg 1$, the rest of them remaining relativistic. Indeed, in such a case we have
\begin{equation}
\mathcal{F}_{\pm}(T)=V(\phi _{\pm})-g_{\pm}\frac{\pi ^{2}}{90}T^{4},
\label{fmaspot}
\end{equation}
where $g_{+}=\sum_{i}c_{i}g_{i}$, with $i$ running over all particle species, while $g_{-}=\sum_{i'}c_{i'}g_{i'} $, where $i'$ runs only over the particles which remain light in the $-$ phase. In most cases of interest
we have $\phi _{+}=0$, and $V(\phi _{+})$ is a constant. In some cases,
$\phi _{-}$ is close to its zero-temperature value, and is also
approximately constant. Hence, we obtain Eq. (\ref{Fbag}), with $\epsilon _{\pm }=V(\phi _{\pm })$ and $a_{\pm }=g_{\pm
}\pi ^{2}/30$. If any of the above conditions is not fulfilled, we expect some deviation from the bag EOS.

\begin{figure}[bth]
\centering
\epsfysize=6cm \leavevmode \epsfbox{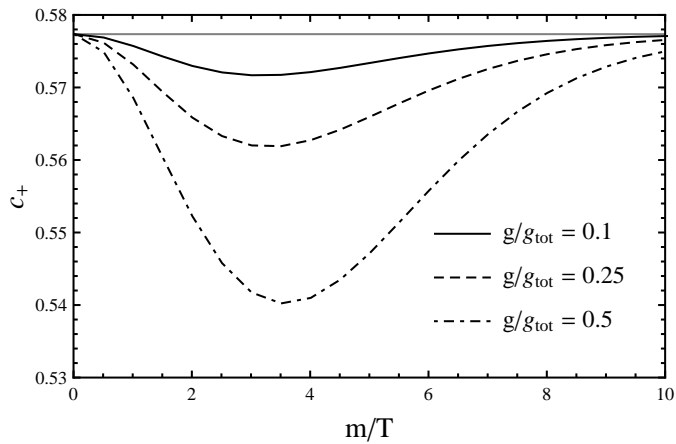}
\caption{The speed of sound for a system with $g_{\mathrm{tot}}$ d.o.f., of which $g$ have a constant mass $m$ and the rest are massless. The horizontal line indicates the value $1/\sqrt{3}$.}
\label{figcmass}
\end{figure}

Since in general we have $\phi_+=0$, the masses $m_i(\phi_+)$ are constant  in the $+$ phase. Thus,
one expects a deviation from Eqs. (\ref{EOSbag}-\ref{Fbag}) only for those particles with $m_i\sim T$ (otherwise the particles either behave like radiation or disappear from the plasma).
A measure of such a deviation  is provided by the dimensionless quantity  $c_+^2=dp_+/de_+=p'_+(T)/e'_+(T)$, where a prime indicates a derivative with respect to $T$.
Assume, for instance, that we have a number $g_{\mathrm{tot}}$ of degrees of freedom, and only $g$ of them have a (constant) mass $m$ in the $+$ phase, the rest of them being massless. We obtain the sound velocity $c_+$ by integrating numerically Eq. (\ref{f1loop2}) as a function of $m/T$. The result is shown in Fig. \ref{figcmass}. The figure corresponds to the case of fermion d.o.f.; the boson contribution is qualitatively and quantitatively very similar. We see that, as expected, for very small or very large values of $m/T$ we obtain the radiation result $c_+=1/\sqrt{3}$. Notice also that the departure from this value is never too large, even if a sizeable fraction of the degrees of freedom have a mass $m\sim T$.

On the other hand, in the low-temperature phase we have $\phi_-\sim T$ (unless the phase transition is very weakly first-order), and some of the masses will fulfill the relation $m(\phi_-)\sim T$.
Moreover, in the general case the value of the minimum $\phi _{-}$ depends on the temperature. This will cause a model-dependent function $c_-(T)$.
In the case $m_i\lesssim T$, we can expand Eq. (\ref{f1loop2}) in powers of $
m_{i}/T$. To quadratic order we have
\begin{equation}
V_{T}(\phi )=\sum_{i}g_{i}\left[ -c_{i}\frac{\pi ^{2}}{90}
T^{4}+\tilde{c}_{i}\frac{m_{i}^{2}}{24}T^{2}\right] ,  \label{vpot}
\end{equation}
where $\tilde{c}_{i}=1$ for bosons and $1/2$ for fermions. Thus, for small  $m_{i}(\phi _{+})/T$ and moderate $m_{i}(\phi _{-})/T$, we have
\begin{eqnarray}
\mathcal{F}_{+}(T)&=&V(\phi _{+})-g_{+}\frac{\pi ^{2}}{90}T^{4}, \\
\mathcal{F}_{-}(T)&=&V(\phi _{-})-g_{+}\frac{\pi ^{2}}{90}T^{4}+bT^{2},
\end{eqnarray}
where $g_{+}=\sum_{i}c_{i}g_{i}$, with $i$ running over all particle species, and $b=\sum_{i'}g_{i'}\tilde{c}_{i'}m_{i'}^{2}(\phi _{-})/24$, where $i'$ runs only over particles which acquire a mass.
Notice that the radiation component is the same in both phases, i.e., the term $\sim T^4$ is proportional to $g_+$ even in the $-$ phase. This is because the massive d.o.f. have not disappeared completely. On the other hand,  we have a correction $\sim T^2$ to the radiation EOS.  If we neglect
the dependence of $\phi _{-}$ on $T$, we have $p_{-}=-\epsilon
_{-}+a_{+}T^{4}/3-bT^{2}$, while in the $+$ phase we may assume $p_{+}=-\epsilon
_{+}+a_{+}T^{4}/3$. In terms of the thermodynamical parameters $T_c$ and $L$ defined above, we have in this case
\begin{equation}
p_{-}(T)=p_+(T)+\frac{L}{2}\left( 1-\frac{T^{2}}{T_{c}^{2}}\right) ,\;\;e_{-}(T)=e_+(T)-\frac{L}{2}\left( 1+\frac{T^{2}}{T_{c}^{2}}
\right) .  \label{hkllm}
\end{equation}
This approximation provides a simple EOS\footnote{A model of this form was already used in Ref. \cite{hkllm}.} with some interesting differences
with respect to the bag EOS. One of them is a  sound velocity $c_{-}(T)\neq 1/\sqrt{3}$.

The model (\ref{hkllm}) can be useful to study the effects of a temperature-dependent speed of sound on the hydrodynamics of phase transition fronts. However, despite the simplicity of
the EOS, the space variation of temperature implies a
space-dependent speed of sound which will make it difficult to avoid a
numerical treatment. We shall address this issue elsewhere. In the present
work we shall neglect the dependence of $c_{\pm }$ on $T$. This
is a reasonable approximation if the
temperature only varies in a small range about $T_{c}$, which is true for
most phase transitions\footnote{Although some
quantities are very sensitive to the departure of $T$ from $T_{c}$ (e.g.,
the bubble nucleation rate or the bubble wall velocity), this is not the
case of the speed of sound.}. In contrast, since
the equation of state may depart significantly form the bag EOS (particularly
in the $-$ phase), the values of $c_{\pm}$ may depart from the bag value $c_s=1/\sqrt{3}$. For the model (\ref{hkllm}) we have
\begin{equation}
c_{-}=\sqrt{\frac{1}{3}}\sqrt{\frac{1-3\alpha }{1-\alpha }},
\end{equation}
where $\alpha =L/(4a_{+}T^{2}T_{c}^{2})$ . We thus see that this EOS gives $c_{-}<1/\sqrt{3}=c_{+}$. For $T=T_{c}$, the parameter $\alpha $ is given by $
\alpha _{c}=L/(4a_{+}T_{c}^{4})=\frac{1}{3}L/w_{+}(T_{c})$. Thus, for instance, for $\alpha _{c}=0.1$ we have $c_{-}(T_c)\simeq 0.51$.

The fact that the equation of state (\ref{hkllm}) can be obtained from the
one-loop free energy through an expansion in powers of $m/T$ rather than in
the limit $m/T\gg 1$, indicates that this model may be more realistic than
the bag in many physical situations.
On the other hand, this EOS has essentially the same number of free
parameters than the bag EOS, and therefore is in principle as limited as the latter in reproducing a general model. In order to
consider general values of the quantities $c_{\pm }$, we will introduce in Sec. \ref{model} a model for which these two quantities are free parameters.

In a general case, the temperature dependence of the minimum $\phi_{-}(T)$ can make the value of $c_-$ depart significantly form that of $c_+$, depending on details of the model.
In order to explore the possible values of the speed of sound in a realistic case, we have considered  the case of the electroweak phase transition,
for a few extensions of the Standard Model (SM). The electroweak phase transition occurs at $T_c\sim 100\mathrm{GeV}$. The number of SM degrees of freedom is $g_*\approx 107$. The Higgs-dependent masses are of the form $m_i=h_i\phi$ and we have  $\phi_+=0$. Thus, in the high-temperature phase the particles are massless. In the low-temperature phase most of the particles remain effectively massless, since the couplings to the Higgs are $h_i\ll 1$ except for the top quark and the $W$ and $Z$ bosons. In the SM the phase transition is weakly first-order, which means that $\phi_-\ll T$, but in extensions of the SM we may have strongly first-order phase transitions.
We have considered models with extra bosons and fermions with masses of the form $m(\phi)=h\phi$ (for details on these models see Refs. \cite{ms10,lms12}).

\begin{figure}[bth]
\centering
\epsfysize=7cm \leavevmode \epsfbox{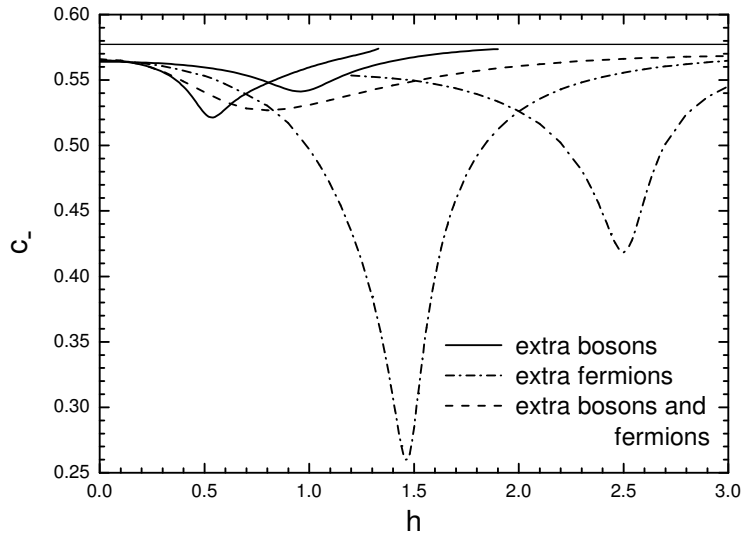}
\caption{The speed of sound in the low-temperature phase of the electroweak phase transition, for some extensions of the Standard Model, as a function of the coupling of the extra particles to the Higgs. The horizontal line indicates the value $c_-=1/\sqrt{3}$.}
\label{figcpt}
\end{figure}

In Fig. \ref{figcpt} we show the value of $c_- (T_c)$ as a function of the coupling $h$. The solid line which is closer to the value  $c_-=1/\sqrt{3}$
corresponds to adding to the SM a scalar field with $g=2$ d.o.f. The other solid line corresponds to a scalar field with $g=12$ d.o.f. The dashed-dotted lines correspond to extra fermions strongly coupled to the Higgs\footnote{This model has also bosons with the same coupling but with higher masses due to $\phi$-independent terms. Hence, these bosons are decoupled from the physics at $T=T_c$ \cite{cmqw05}.}. Again, we have considered $g=2$ and $g=12$ extra d.o.f., and the case $g=2$ is that with the smaller departure from the radiation case. We have also considered a case of extra bosons and fermions (dashed line) with the same coupling $h$ and the same number of d.o.f., $g=12$.

It is interesting to check that the speed of sound approaches the relativistic value in the two expected limits; namely, for small $h$ and large $h$.
Indeed, as $h\to 0$ the mass of the  extra particle vanishes and the species behaves as radiation. The exact value $c_-=1/\sqrt{3}$ is not reached in this limit since we still have the massive SM particles. For large $h$, the extra particles become very massive and decouple from the plasma. Besides, in some cases the phase transition becomes very strong, with large values of $\phi/T$, and even the SM particles acquire large values of $m/T$ (essentially, because $T_c$ becomes small). This is why the solid lines in Fig. \ref{figcpt} quickly approach the value $c_-=1/\sqrt{3}$ as $h$ is increased. Indeed, for larger values of $h$ than those considered in the figure, the phase transition becomes too strong and these models become unphysical.

We observe that the departure from the radiation EOS is more significant for the case of strongly coupled fermions. Nevertheless, notice that even with only two extra bosonic d.o.f. we may have a departure from the radiation value which is comparable to the strongest of the cases considered in Fig. \ref{figcmass} for a constant mass.
Such a difference between the possible values of $c_-$ and $c_+$ is due to  the temperature dependence of the minimum $\phi_{-}(T)$. This induces a temperature dependent mass $m(\phi_-(T))$ as well as a temperature dependent ``vacuum'' energy density $V(\phi_{-}(T))$. Both affect the value of $c_-^2=p'_-(T)/e'_-(T)$.

\begin{figure}[bth]
\centering
\epsfysize=6cm \leavevmode \epsfbox{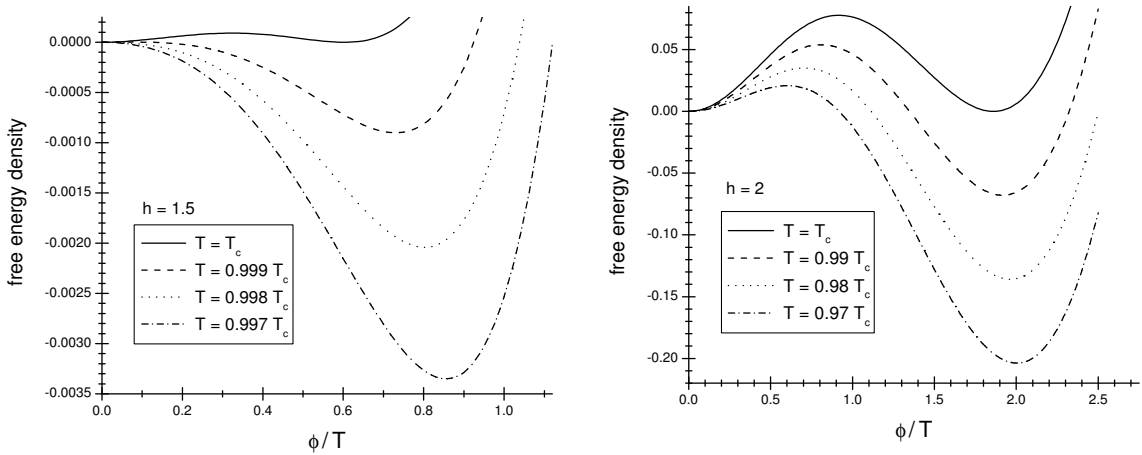}
\caption{Plot of $\mathcal{F}(\phi,T)/T^4-\mathcal{F}(0,T)/T^4$ as a function of $\phi/T$ for the extension of the SM with 12 strongly coupled fermions, for $h=1.5$ (left panel) and $h=2$ (right panel).}
\label{figpots}
\end{figure}

Notice that the  light SM d.o.f. behave like radiation, and their contribution to the free energy density is of the order of $100T^4$. On the other hand, the contribution of the particles with temperature-dependent masses is smaller since it is proportional to their number of d.o.f., $g$. On dimensional grounds, this contribution is of order $gT_c^4$. Similarly, the ``vacuum'' part is of order $T_c^4$. Since we have considered values of $g\lesssim 10$, the radiation part of the free energy density should be a factor of $10$ higher than the non-radiation parts, and one may wonder why the deviations from $c_-^2=1/3$ are so large in some cases. In fact, the speed of sound involves derivatives of these contributions. Thus, the light particles contribute terms of order $400T_c^3$ while the particles which become heavy contribute terms $\sim g(d\phi_-/dT)T_c^3$. The latter derivative is in many cases large at $T=T_c$, and in some cases we even have $d\phi_-/dT\sim 100$. As a consequence, the latter contribution may be larger than the light particles contribution\footnote{More precisely, the terms proportional to  $d\phi_-/dT$ will cancel out in the derivative $dp_-/dT$ and will increase the derivative $de_-/dT$, causing a lower value of $c_-^2$. Indeed, notice that $\mathcal{F}_-(T)$ is given by $\mathcal{F}(\phi,T)$ evaluated at $\phi=\phi_-$. Then we have $s_-=dp_-/dT=-d\mathcal{F}_-/dT=-(\partial \mathcal{F}/\partial\phi )( d\phi_-/dT)-\partial\mathcal{F}/\partial T$. Since $\phi_-$ is the minimum of $\mathcal{F}$, we obtain $s_-=-\partial\mathcal{F}/\partial T$ [it is easy to check, from the two plots shown in Fig. \ref{figpots}, that the size of the variation of $\mathcal{F}_-$ with $T$ is indeed uncorrelated to that of $\phi_-$]. On the other hand, for $de_-/dT=Tds_-/dT$, the terms proportional to $g d\phi_-/dT$ (and dimensionally $\sim T^3$) will not cancel out but will give a positive contribution.}. To show this effect, we consider in Fig. \ref{figpots} the plot of the difference $\Delta \mathcal{F}(\phi,T)= \mathcal{F}(\phi,T)-\mathcal{F}(0,T)$ (normalized to $T^4$), i.e., we subtracted the radiation contribution from the free energy density. The left panel corresponds to the case of 12 strongly coupled extra fermions with $h=1.5$ (which, according to Fig. \ref{figcpt}, has a large departure from the radiation case). We see that, indeed, the low-temperature minimum has a variation $\delta\phi_-\sim 0.1T_c$ in a temperature range  $\delta T\sim 10^{-3}T_c$ (hence, $d\phi_-/dT\sim 100$). For comparison we also show the case of $h=2$ (right panel), corresponding to a stronger phase transition. In this case the minimum $\phi_-$ is closer to its zero-temperature value and varies very little in a larger temperature range. In such a case we have a smaller deviation from $c_-^2=1/3$.

We thus conclude that in a realistic case we will most likely have $c_-<c_+\simeq 1/\sqrt{3}$. In particular, the one-loop effective potential gives the sound velocity bound  $c_s<1/\sqrt{3}$. This seems to be a general bound for the speed of sound in any medium, including strongly coupled field theories like QCD \cite{bound}, although there is no fundamental reason for this bound \cite{unbound}. Regarding the relation $c_-<c_+$, although it is the most probable case,
we may also have $c_-=c_+$ or even $c_->c_+$.
For instance, consider a particle species which has a mass given by $m^2(\phi)=m_0^2+h^2\phi^2$. For $m_0\sim T$ and $h$ large enough we may have $m(\phi_+)\sim T$, $m(\phi_-)\gg T$. The contribution of this species alone to the effective potential can make the phase transition strongly first-order. On the other hand,  in the $+$ phase this contribution will tend to lower the speed of sound from its radiation value, while in the $-$ phase the particle will disappear from the plasma, which may thus behave as radiation. In such a case we will have $c_+< 1/\sqrt{3}$, $c_-\simeq 1/\sqrt{3}$.

The bag EOS has been often used as an approximation for realistic models.
However, there is some ambiguity in the way to choose the bag free parameters for such an approximation.
The computations using the bag EOS generally depend on the false-vacuum energy-density difference $\Delta \epsilon=\epsilon_+-\epsilon_-$.
For a general model, though, the quantity $\Delta \epsilon$ has no clear meaning at finite temperature. In particular, as already discussed, the quantity $\epsilon _{-}(T)=V(\phi _{-}(T))$ does
not generally behave like vacuum energy density.
With no loss of generality, we may write the total energy density in the
form $e_{\pm }(T)=a_{\pm }(T)T^{4}+\epsilon _{\pm }(T)$. However, such a
separation is meaningless unless there is some criterion to choose the functions $a_{\pm
}(T) $ and $\epsilon _{\pm }(T)$. Requiring that these functions provide a decomposition of the pressure in bag form, $p_{\pm }(T)=a_{\pm }(T)T^{4}/3-\epsilon _{\pm }(T)$, as well as that of $e_\pm$ above, one obtains \cite{ekns10}
\begin{equation}
a_{\pm }=3w_{\pm }/(4T^{4}),\quad \epsilon _{\pm }=
(e_{\pm }-3p_{\pm })/4.  \label{fitekns}
\end{equation}
These equations give an unambiguous definition of $a_{\pm
}(T) $ and $\epsilon _{\pm }(T)$. However, the utility of this approach is not clear. In order to actually use the bag EOS, we need constant parameters. We may thus use, e.g. the values $a_{\pm }(T_c)$ and $\epsilon _{\pm }(T_c)$. For the bag to be a good
approximation, the quantities $a_{\pm }(T)$ and $\epsilon _{\pm }(T)$
should be approximately constant.

Therefore, one may write the few bag parameters in terms
of thermodynamic quantities which are well defined in any model. Since there is no single way of doing this, one may choose the physical quantities which are more relevant for a given calculation (rather than $w_\pm$ and $e_\pm-3p_\pm$). Several calculations using the bag EOS depend on the single variable $\alpha=\Delta\epsilon/(a_+T^4)$.
This can be written (for the bag EOS) in the form $\alpha=L/(4a_+T^4)$. The latent heat $L$ is well defined in any model. The same is often true for the radiation energy density $e_R=a_+T^4$ in the $+$ phase, but sometimes the temperature dependence is not of this form. For a general model, we may define a ``thermal'' energy density $e_R(T)=e_+(T)-\epsilon_+$ \cite{ms10,lms12}, provided that the vacuum energy density $\epsilon_+=V(\phi_+)$ is a constant for the $+$ phase. Thus, writing $\alpha=L/(4e_R)$, we have a means to calculate this variable for any model. A better solution would be  to express $\alpha$ in the form $\alpha=(T_c/T)^4\alpha_c$, with $\alpha_c=\frac{1}{3}L/w_{+}(T_{c})$.
The parameters $L$ and $w_+(T_c)$ are clearly defined in any model, as well as the amount of supercooling $T/T_c$.
If the given model is well approximated by the bag EOS, all these approaches should not give significant differences.

\section{Hydrodynamics \label{hidrogral}}

The propagation of a phase transition front in the plasma causes reheating as well as bulk motions of the fluid. We will be mostly interested in these effects and not in the backreaction of the fluid disturbances on the wall.

\subsection{Fluid equations}

We will consider a bubble wall propagating with a constant velocity $v_{w}$.
Away from the wall, which we assume to be infinitely thin, the plasma is a
relativistic fluid characterized by its energy-momentum tensor
\begin{equation}
T^{\mu \nu }=\left( e+p\right) u^{\mu }u^{\nu }-pg^{\mu \nu },  \label{tmunu}
\end{equation}
where $e$ and $p$ are the energy density and pressure in the local rest
frame of the fluid element \cite{landau}, and $u^{\mu }=\gamma (1,\mathbf{v})$,
with $\gamma =1/\sqrt{1-v^{2}}$, is the four-velocity field. The fluid
equations are obtained from the conservation of $T^{\mu \nu }$,
\begin{equation}
\partial _{\mu }T^{\mu \nu }=0.  \label{consT}
\end{equation}

The enthalpy and other quantities are discontinuous at the interface, and so
will be the fluid velocity and temperature. The relations between variables
in front and behind the wall (which we denote with subindexes $+$ and  $-$,
respectively) are obtained by integrating Eqs. (\ref{consT}) across the
interface.
For simplicity, we shall consider a planar wall moving towards the
positive $z$ direction, and a fluid velocity perpendicular to the wall.
Hence, in the reference frame of the wall we have $v_{x}=v_{y}=0$,
and $v_{z}\equiv -v$. As a consequence, the fluid profile will have plane
symmetry. The generalization to other wall geometries is straightforward
\cite{k85,lm11}.
Thus, we have, in the wall frame, an incoming flow with velocity $-v_{+}$ and an outgoing flow with velocity $-v_{-}$, which are related by \cite{landau}
\begin{eqnarray}
w_{-}v_{-}^{2}\gamma _{-}^{2}+p_{-} &=&w_{+}v_{+}^{2}\gamma _{+}^{2}+p_{+},
\label{land1} \\
w_{-}v_{-}\gamma _{-}^{2} &=&w_{+}v_{+}\gamma _{+}^{2}.  \label{land2}
\end{eqnarray}
These equations are local and therefore are independent of the bubble shape, as long as $v_\pm$ represents the
component of the fluid velocity perpendicular to the interface\footnote{The equation for the components parallel to the wall are simply the continuity conditions for these components. The case of non-vanishing $v_{x},v_{y}$ is
important when considering the hydrodynamic stability of the stationary motion \cite{hkllm,landau,mm14,link,instdeto}.}.

Eqs. (\ref{land1}-\ref{land2}), together with the EOS, determine $v_{+}$ as a
function of $v_{-}$ and $T_{+}$. For a given $T_{+}$ we have in general two
solutions for $v_{+}$ vs. $v_{-}$ (we shall consider concrete examples in Sec. \ref{model}). Thus, the propagation of the front is classified according to these two kinds of solutions, called detonations and deflagrations. For detonations we have $v_{+}>v_{-}$. In the range $0<v_{-}<1$, the value
of $v_{+}$ has a minimum at $v_{-}=c_{-}$, which is called the \emph{Jouguet
point}. The minimum value of $v_{+}$ is the Jouguet velocity
$v_{J}^{\mathrm{\det }}$. The condition $v_{+}>v_{-}$ implies
$v_{J}^{\mathrm{\det }}>c_{-}$. For the bag EOS this implies also
the condition $v_{+}>c_{+}$. For deflagrations, in contrast, we have $v_{+}<v_{-}$, and $v_{+}$ has a maximum value $v_{J}^{
\mathrm{def}}$ at the Jouguet point $v_{-}=c_{-}$. We thus have $v_{J}^{\mathrm{def}}<c_{-}$, which for the bag case is equivalent to $v_{+}<c_{+}$.
These hydrodynamic propagation modes are further divided into \emph{weak}
and \emph{strong} solutions. The solution is called weak if the velocities $
v_{+}$ and $v_{-}$ are either both supersonic or both subsonic. Otherwise,
the solution is called strong. It turns out that strong solutions are not
possible in a cosmological phase transition (see Sec. \ref{model}).

Due to the motions of the fluid, discontinuities in the same phase may also arise, which are called \emph{shock fronts} \cite{landau}. In the reference frame of a shock front, Eqs. (\ref{land1}-\ref{land2}) apply, only that we have the same EOS on both sides of the discontinuity. These equations give in particular the velocity of the shock front.

The equations for the fluid away from these discontinuities can also be derived from Eqs. (\ref{consT}), taking into account that the energy density and the pressure in Eq. (\ref{tmunu}) are further related by the EOS.
The fluid equations can be written in terms of the
dimensionless quantity $c_s^{2}=dp/de$ (see, e.g., \cite{k85,lm11}). Furthermore, the equations are considerably simplified by assuming the similarity condition \cite{landau}, namely, that the fluid velocity and thermodynamical variables depend only on the ratio $\xi =z/t$, where $t$ the time since bubble nucleation. This is justified because there is no characteristic distance scale in the equations. The similarity condition implies that any fixed point in the fluid profile moves with constant velocity $\xi $. This condition is compatible with a bubble wall moving with constant velocity.
Thus, for a wall at position  $z_{w}=v_{w}t$ and a shock front at position  $z_{\mathrm{sh}}=v_{\mathrm{sh}}t$, the fluid profile will have discontinuities at $\xi _{w}=v_{w}$ and $\xi _{\mathrm{sh}}=v_{\mathrm{sh}}$.

For the planar case, the fluid equations give very simple solutions (see,
e.g., \cite{s82,k85,lm11}),
\begin{eqnarray}
v(\xi ) &=&\mathrm{constant},  \label{vconst} \\
v_{\mathrm{rar}}(\xi ) &=&\frac{\xi -c_s}{1-c_s\xi }.  \label{vrar}
\end{eqnarray}
For the solution (\ref{vconst}) we have an arbitrary constant which must be determined by boundary or matching conditions. In
contrast, the  solution (\ref{vrar}) is fixed. It is a monotonically increasing function of $\xi$ which vanishes at $\xi =c_s$. Physically, this solution corresponds to a rarefaction wave.
The solution $v_{\mathrm{rar}}$ is valid in any inertial reference frame. We
shall be interested in fluid profiles in the reference frame of the bubble
center, where we have positive fluid velocities\footnote{In contrast, in the reference frame of the wall we have negative fluid velocities, and a negative part of the solution must be used \cite{mm14}. There is another solution which we are not considering here, namely, $v(\xi )=(\xi +c)/(1+c\xi)$. This solution may contribute to the fluid profile for a front propagating in the opposite direction (i.e., for values of $\xi<0$).} and this solution is only possible for $\xi >c_s$.
For the solutions $v=$ constant we have constant enthalpy, while for the solution $v_{\mathrm{rar}}$ we have \cite{landau,s82,k85}
\begin{equation}
w=w_{0}\exp \left[ \int_{\xi _{0}}^{\xi }\left( \frac{1}{c_s^{2}}+1\right)
c_s\gamma ^{2}\frac{dv}{d\xi }d\xi \right] ,  \label{enth}
\end{equation}
where $w_{0}=w(\xi _{0})$. From this equation we may obtain the temperature, since the EOS gives $w=w(T)$ and $c_s=c_s(T)$. Thus, in the general case we may insert Eq. (\ref{vrar}) in Eq. (\ref{enth}) to obtain an equation for $T(\xi )$. Then, from $T(\xi )$ we readily obtain $c_s(\xi )$ from the EOS and, inserting $c_s(\xi )$ in Eq. (\ref{vrar}), we finally obtain $v(\xi )$. The case of constant $c_s$ is much simpler, since Eq. (\ref{vrar}) already gives the velocity profile while Eq. (\ref{enth}) gives $w(v)$.

The fluid profiles must be constructed from these solutions, using the
boundary conditions and the matching conditions (\ref{land1}-\ref{land2}).
The boundary conditions are that the fluid velocity vanishes at $\xi =0$
(i.e., at the center of the bubble) and at $\xi =1$ (i.e., far in front of
the wall). There is also a boundary condition for $T$, namely, that the
temperature far in front of the wall is that at which the bubble nucleated, $
T=T_{n}$.
For the bag EOS there are three kinds of solutions, which we briefly review here (for an exhaustive construction of these solutions see Ref. \cite{lm11}).
In Sec. \ref{model} we shall consider all the possible fluid profiles for a model which generalizes the bag EOS.

For the bag EOS, the Jouguet point $v_-=c_-=c_+$ separates weak from strong solutions. Thus, for weak or Jouguet detonations we have $v_+>v_-\geq c_-$, and the incoming flow (in the wall frame) is always supersonic. As a consequence, the fluid in front of the wall is unperturbed by the latter. This means that the fluid velocity vanishes there and the temperature is given by the nucleation temperature. Hence, the wall is supersonic ($v_w=v_+\geq v^{\mathrm{det}}_J$).

On the other hand, for weak or Jouguet deflagrations we have $v_+<v_-\leq c_-$. The incoming flow is thus subsonic, and we have a shock wave propagating in front of the wall. Relative to the fluid behind it, the wall moves at a velocity $v_-\leq c_-$. Nevertheless, such a wall may still be supersonic if the fluid behind it also moves with respect to the bubble center. Indeed, in such a case the wall velocity $
v_{w} $ is the relativistic sum of $v_{-}$ and the velocity $\tilde{v}_{-}$ of the fluid with respect to the bubble center. As a consequence, we have two kinds of deflagration solutions: a subsonic wall with velocity $v_w=v_-<c_-$, which is a weak deflagration, and a supersonic wall with $v_-=c_-$ and a rarefaction wave following it, which is a Jouguet deflagration.
The supersonic Jouguet deflagration turns out to fill the
range  $c_{-}\leq v_{w}\leq v_{J}^{\det }$.

It is evident that the discussion above (and, hence, the properties of the profiles) may suffer modifications in the case $c_{+}\neq
c_{-}$. For instance, for $c_{+}>c_{-}$ we could in principle have a
weak detonation ($v_{+}>v_{-}\geq c_{-}$) which is subsonic with respect to
the fluid in front of it ($v_{+}<c_{+}$). In such a case, one expects that a
supersonic shock front will propagate in front of the detonation front. We
shall discuss this possibility in the context of a specific model in Sec.
\ref{model}.

\subsection{Distribution of the released energy}

The energy density of the fluid is given by the $00$ component of Eq. (\ref{tmunu}). In the present case it can be split in several forms,
\begin{equation}
T^{00}(v)=w\gamma ^{2}-p=\left(
e+pv^{2}\right) \gamma ^{2}=e+wv^{2}\gamma
^{2}. \label{T00}
\end{equation}
The quantity appearing in the last term of the last member of Eq. (\ref{T00}),
\begin{equation}
e_v=w\gamma^2v^2,\label{defkin}
\end{equation}
turns out to be relevant for the calculation of gravitational waves. Indeed, the gravitational radiation depends on the volume integral of this quantity  \cite{kkt94}.
Since we have $T^{00}(v= 0) =e$, we may write
\begin{equation}
e_v=T^{00}\left( v\right) -T^{00}\left( 0\right) \label{defkin2}.
\end{equation}
As a consequence,
it is usual to associate this quantity to the kinetic energy density in macroscopic motions
of the fluid (see, for instance, \cite{kkt94,ekns10,lm11,kmk02}).
It is worth mentioning that, in the non-relativistic limit (i.e., for a
non-relativistic gas and small $v$), the quantity $e_v$
becomes $\rho v^{2}$ instead of the expected result $\frac{1}{2}\rho
v^{2}$. On the other hand, it is well known \cite{landau} that the total energy
density $T^{00}(v)$ does give the expected limit $\rho +\frac{1}{2}\rho
v^{2}$, where $\rho $ is the mass density in the ``laboratory'' frame\footnote{In the non-relativistic limit, we have $p\to 0$ and $T^{00}\to e\gamma^2$, but $e$ is the \emph{proper} energy density, and we have $e\to\rho_p=\rho/\gamma$. This gives $T^{00}(v)\to \rho\gamma\to\rho(1+\frac12 v^2)$. The same reasoning gives, for either of Eqs. (\ref{defkin}-\ref{defkin2}) [taking into account that $T^{00}(0)\to\rho_p$], $e_v\to \rho_pv^2= \rho v^2+\mathcal{O}(v^4)$.}. This
discrepancy is due to the fact that $e_v$ is given by the \emph{difference} of
two energy \emph{densities}. Consider a given fluid element which has a volume
$V(0)$ when it is at rest and a volume  $V(v)$ when it moves with velocity $v$. If we define the kinetic energy of this fluid element as  $E_{\mathrm{kin}}=E(v)-E(0)=T^{00}(v)V(v)-T^{00}(0)V(0)$, then we could define the kinetic energy density as ${e}_{\mathrm{kin}}=E_{\mathrm{kin}}/V(v)=T^{00}(v)-\gamma T^{00}(0)$,
which gives the correct non-relativistic limit\footnote{In Ref. \cite{cd06}, the definition $e_{\mathrm{kin}}=(1/2)wv^{2}\gamma ^{2}$ is
used, which also gives the correct limit and is proportional to $e_v$.}. However, such a definition of a density involving two different volumes is not satisfactory either.
Since Eq. (\ref{T00}) provides a natural splitting of the total energy density into a velocity-independent part $e(T)$ and a part which vanishes for $v=0$, namely, the quantity $e_v$, in the relativistic context this quantity is more appropriate for quantifying the macroscopic kinetic energy of the fluid.

In any case, it is the quantity $e_v$ the one which enters the calculation of gravitational waves.
For the bag model, it is customary to define the \emph{efficiency factor}
\cite{kkt94}
\begin{equation}
\kappa =E_v/\left( \Delta \epsilon V_{b}\right) ,
\label{kappadef}
\end{equation}
where $E_v$ is the space integral of Eq. (\ref{defkin}) and $V_{b}$ is the volume of the bubble. For a constant wall
velocity and a fluid which satisfies the similarity condition, the numerator
and the denominator in Eq. (\ref{kappadef}) have the same behavior with
time. As a consequence, $\kappa $ does not depend on time. For instance,
in the planar wall approximation the bubble walls are at a distance $
z_{w}=v_{w}t$ from a symmetry plane. The volume of the bubble is thus
proportional to $z_{w}$, and the kinetic  energy $E_v$ is proportional $\int_{0}^{\infty }wv^{2}\gamma ^{2}dz$. Taking into account that $w$ and $v $ depend only on $\xi =z/t$, we have
\begin{equation}
\kappa =\frac{1}{\Delta \epsilon v_{w}}\int_{0}^{\infty }wv^{2}\gamma
^{2}d\xi .  \label{kappa}
\end{equation}

We wish to point out a
wrong interpretation of
the factor $\kappa $ (which does not necessarily
lead to incorrect results, as long as the quantity $\kappa \Delta
\epsilon $ is correctly used for the calculation of gravitational waves).
According to Eq. (\ref{kappadef}), the factor $\kappa $ gives the ratio of
the kinetic energy to the released vacuum energy. The vacuum energy is often confused with the total available energy or latent heat (see, e.g., Ref. \cite{kkt94}). As a consequence, $\kappa$ is generally
interpreted as the fraction of the released energy which goes into bulk
kinetic energy, and a fraction $1-\kappa $ is supposed to go into thermal energy (see, e.g., \cite{ekns10}). However, $\Delta \epsilon V_{b}$ is not the
total energy released in the phase transition, since the internal energy of
the system is not comprised of vacuum energy alone. For the bag EOS, the
latent heat $L\equiv e_{+}(T_{c})-e_{-}(T_{c})$ is given by $L=4\Delta
\epsilon $; i.e., the released energy at $T=T_c$ is quite larger than $\Delta
\epsilon V_b$. Hence, the proportion of energy which goes into increasing the thermal energy will be larger than $1-\kappa$.

A more appropriate definition of the efficiency factor
would thus be ${\kappa}=E_{v}/( LV_{b}) $. However, the phase transition does not occur exactly at $T=T_c$ since there is always a certain amount of supercooling.
At a temperature $T_{n}<T_{c}$, the energy difference between the two phases is given by
\begin{equation}
\Delta e_{n}=e_{+}( T_{n}) -e_{-}( T_{n})  \label{den}
\end{equation}
rather than by $L= e_{+}(T_{c})-e_{-}(T_{c})$.
Although  the temperature $T_n$ is in general very close to $T_c$, for an extremely supercooled phase transition occurring at $T_n=0$ we would
have $\Delta e_n=\Delta \epsilon$.
For a bubble expanding at $T=T_n$, we may assume that the released energy is given by $\Delta e_{n}V_{b}$.

In practice, the temperature in the $-$ phase will never be given by the temperature $T_n$, since the energy that is released during the phase conversion at the bubble walls will cause reheating (in general, both inside and outside the bubble) as well as fluid motions. In a stationary state we have stationary profiles for the fluid temperature and velocity. Thus, as the bubble volume changes by $\delta V_b$, the energy difference $\Delta e_n\delta V_b$ goes instantaneously into maintaining the fluid profile.
We will now show that the released energy $\Delta e_n V_b$ naturally splits into the energy $E_v$ and a quantity $\Delta E_r$ which we may interpret as the energy used to reheat the plasma.

Consider energy conservation in a volume $V$ which includes the bubble and the region of the
fluid which is being perturbed. The energy in this volume is
given by the integral of Eq. (\ref{T00}), $E_{f}=\int_{V}e+E_{v}$, while initially it was given by $E_{i}=e_{+}\left( T_{n}\right) V$.
Neglecting for simplicity the loss of energy due to
the adiabatic expansion, we have $E_{f}=E_{i}$, which gives
\begin{equation}
E_{v}+\int_{V}\left[ e( T) -e_{+}( T_{n})
\right] =0.  \label{balance}
\end{equation}
According to this equation, the integral $\int_{V}\left[ e( T) -e_{+}( T_{n})
\right] $ is negative and cannot be interpreted as the energy which goes into reheating. Since the EOS is different in the two phases, it is convenient to separate this integral,
\begin{equation}
E_v+\int_{V_{b}}\left[
e_{-}( T) -e_{+}( T_{n}) \right] +\int_{\bar{V}_{b}}
\left[ e_{+}( T) -e_{+}( T_{n}) \right] =0. \label{balancesep}
\end{equation}
where $\bar{V}_{b}=V-V_{b}$. The last integral in Eq. (\ref{balancesep}) quantifies the energy used to change the temperature from $T_n$ to $T(\mathbf{x})$ in the $+$ phase. On the other hand, the first integral cannot be interpreted in this way, since it involves a change of phase.
Subtracting
and adding $e_{-}\left( T_{n}\right) $, we decompose this integral as
$-\Delta e_{n}V_{b}+\int_{V_{b}}\left[ e_{-}\left( T\right) -e_{-}\left(
T_{n}\right) \right]$. The first of these two terms may be interpreted as the energy which the system has to loose to change phase at $T=T_n$, while the second quantifies the energy used to reheat the system from $T_n$ to $T(\mathbf{x})$. Thus, Eq. (\ref{balancesep}) becomes
\begin{equation}
E_{v}+\Delta E_{r}=\Delta e_{n}V_{b}, \label{distr}
\end{equation}
where
\begin{eqnarray}
\Delta E_{r}&=&\int_{V_{b}}\left[ e_{-}( T) -e_{-}(
T_{n}) \right] +\int_{\bar{V}_{b}}\left[ e_{+}( T)
-e_{+}( T_{n}) \right]   \label{Der} \\
&=&\int_{V}\left[ e( T) -e(
T_{n}) \right].   \label{Der2}
\end{eqnarray}

Taking into account Eq. (\ref{distr}), we define a new efficiency factor
\begin{equation}
\tilde{\kappa}=\frac{E_{v}}{\Delta e_{n}V_{b}},
\label{kappatilde}
\end{equation}
which quantifies the fraction of the released
energy which goes into bulk motions.  From Eq. (\ref{distr}) we have
\begin{equation}
1-\tilde{\kappa}=\frac{\Delta E_{r}}{\Delta e_{n}V_{b}}.  \label{fractherm}
\end{equation}
which we interpret as the fraction of the released energy which goes into
reheating\footnote{According to the discussion below Eq. (\ref{defkin2}), an alternative kinetic energy density ${e}_{\mathrm{kin}}=wv^2\gamma^2+e-\gamma e$ could be used, since the energy density $e$ corresponds to the local rest frame of the fluid element. Similarly, we could use $\gamma e(T)$ instead of $e(T)$ in Eq. (\ref{Der2}). This gives an alternative definition of $\Delta E_r$ which, together with $\int e_{\mathrm{kin}}$, give an alternative splitting of $\Delta e_n V_b$ to that of Eq. (\ref{distr}).}.
For planar walls we have, in analogy to Eq. (\ref{kappa}),
\begin{equation}
\tilde{\kappa} =\frac{1}{\Delta e_{n}v_{w}}\int_{0}^{\infty }wv^{2}\gamma
^{2}d\xi .  \label{kappatil}
\end{equation}
Notice that $\tilde{\kappa}$ is model independent, since the quantity $\Delta e_n$ is defined in any model, in contrast to the bag parameter $\Delta \epsilon$.
For the bag EOS and for $T_{n}$ close to $T_{c}$, we have the relation $\tilde{\kappa}\simeq \kappa /4$ (and $\Delta e_{n}\simeq L$).

A last comment on the interpretation of $\kappa$ is worth. For the bag EOS,  Eq. (\ref{balance}) may be rewritten in the form
\begin{equation}
E_{v}+\int_{V}\frac{3}{4}\left[ w( T) -w_{+}(
T_{n}) \right] =\Delta \epsilon V_{b},  \label{frackappa}
\end{equation}
which  leads to \cite{ekns10}
\begin{equation}
1-\kappa =\frac{1}{\Delta \epsilon V_{b}}\int_{V}\frac{3}{4}\left[
w(T)-w_{+}( T_{n})\right] .  \label{wrongkappa}
\end{equation}
According to our discussion above, the integral in Eqs. (\ref{frackappa}-\ref{wrongkappa}) should not be interpreted as the reheating energy. Indeed, according to our definition (\ref{Der}), the latter can be written as $\Delta E_{r}=\int_{V}\frac{3}{4}\left[ w( T)
-w( T_{n}) \right] $ (notice the absence of a $+$ index in this expression). We may thus write
\begin{equation}
1-\tilde{\kappa}=\frac{1}{\Delta e_{n}V_{b}} \int_{V}\frac{3}{4}\left[ w(T)
-w(T_{n}) \right].  \label{rightkappa}
\end{equation}
In Ref. \cite{ekns10} the integral
in Eq. (\ref{wrongkappa}) is written as $\int \frac{3}{4}\left[
w(T)-w_{+}(T_{n})\right]=\int \left[ e(T)-e_{+}(T_n)\right] $, which is then interpreted as the energy used to increase the thermal energy. However, this last equality is wrong and, in any case,
we have, according to
Eq. (\ref{balance}), $\int \left[ e(T)-e_{+}(T_n)\right] =-E_v<0$. We remark that  Eqs. (\ref{wrongkappa}) and (\ref{rightkappa}) are equivalent. The differences are in the way in which the energy is arranged and in the interpretation. Outside the bubble, the integrands in Eqs. (\ref{wrongkappa}) and (\ref{rightkappa}) coincide. Inside the bubble, the integrand in Eq. (\ref{wrongkappa}) is given by $\frac{3}{4}\left[
w_-(T)-w_{+}(T_{n})\right]$, while in Eq. (\ref{rightkappa}) we have $\frac{3}{4}\left[ w_-(T)-w_-(T_{n}) \right]$. We interpret the latter as the energy density involved in the change of temperature from $T_n$ to $T$ in the $-$ phase, and the whole integral in Eq. (\ref{rightkappa}) as the part of the total released energy $\Delta e_{n}V_{b}$ which goes into reheating.

Apart from these interpretation issues, the quantity $\kappa$ obtained using the bag EOS  is useful (e.g., for the calculation of gravitational waves) because it is relatively easy to calculate as a function of the quantities $v_{w}$ and $\alpha _{n}=\Delta \epsilon /(a_{+}T_{n}^{4})$
(see the next section).
However, as discussed in previous sections, the
quantity $\Delta \epsilon $ is not clearly defined in models other than
the bag, and care must be taken in applying the result for $\kappa
(v_{w},\alpha _{n})$.
On the other hand,
the definition (\ref{kappatilde})  is as useful as Eq. (\ref{kappadef}) for applications. Indeed, the quantity $E_v$ can be obtained either from $\kappa \Delta \epsilon V_b$ or from $\tilde{\kappa}\Delta e_n V_b$. For the bag EOS, the relation between these quantities is given by $\kappa/\tilde \kappa=\Delta e_n/\Delta \epsilon=1+3T_n^4/T_c^4$.
In the next section we shall calculate analytically $\tilde{\kappa}$ and $\Delta e_n$
for for the generalized version of the bag EOS.

\section{A generalization of the bag EOS \label{model}}

We wish to take into account the possibility that the speed of
sound has arbitrary values $c_{+}$ and $c_{-}$ in each phase. However, for simplicity, we want to consider the case of constant $c_{\pm}$. This condition gives a model which is almost as simple as the
bag EOS, but has an extra free parameter for each phase.

\subsection{The model}

The condition $\partial p/\partial e=c_s^{2}=$ \emph{constant} restricts the
equation of state to the form $p=c_s^{2}e+$ \emph{constant}. Using the
relation $e+p=Ts$ with $s=dp/dT$, we find that $e$ and $p$ are necessarily
given by
\begin{equation}
e=aT^{\nu }+\epsilon ,\quad p=c_s^{2}aT^{\nu }-\epsilon ,
\label{EOSgral}
\end{equation}
where $a$ and $\epsilon $ are constants, and the exponent $\nu $ is given
by
\begin{equation}
\nu =1+1/c_s^{2}.
\end{equation}
This is the most general EOS with a constant speed of sound.

A system described by this EOS has two components: a \textquotedblleft
vacuum\textquotedblright\ energy density (the constant $\epsilon $) and a
\textquotedblleft thermal\textquotedblright\ energy density (the temperature
dependent part $aT^{\nu }$). The exponent $\nu $ can take any value between $
2$ (corresponding to $c_s=1$) and $\infty $ (corresponding to $c_s=0$). For $
c_s^{2}=1/3$ we have $e=aT^{4}+\epsilon $, i.e., vacuum energy plus
radiation. For $c_s^{2}\neq 1/3$, the coefficient $a$ is dimensional.
We could try to give an interpretation to this  EOS by
considering an effective, temperature-dependent radiation constant $a(T)=aT^{\nu-4}$. Such an interpretation might be useful if it provided a way to choose the value of $\nu$ so that $a(T)$ behaves in some physical manner. However, the general behavior of such an effective $a(T)$ is not clear a priori, as discussed in Sec.
\ref{EOSs}.
As in the bag case, we may set the parameters so that they give the required values of physical quantities. In this case we can also set the speed of sound.

We wish to use equations of state of the form (\ref{EOSgral}) to describe a
phase transition. Therefore, we consider two phases. The bag EOS will be a particular case of this model, corresponding to $\nu =4$ in
both phases. For the moment, let us denote these phases with the indices 1 and 2. The system is thus described by
\begin{equation}
\begin{array}{ccc}
e_{1}=a_{1}T^{\nu _{1}}+\epsilon _{1}, &  & e_{2}=a_{2}T^{\nu
_{2}}+\epsilon _{2}, \\
p_{1}=c_{1}^{2}a_{1}T^{\nu _{1}}-\epsilon _{1}, &  &
p_{2}=c_{2}^{2}a_{2}T^{\nu _{2}}-\epsilon _{2.}
\end{array}
\label{EOS}
\end{equation}
The quantities $\epsilon _{1}$ and $\epsilon _{2}$ give
the vacuum energy density of each phase.
At very low temperatures, the stable phase will be the one with the lower
value of $\epsilon $, whereas at very high temperatures it will be that
with the largest exponent $\nu $. However, at intermediate temperatures, we may have different situations, depending on the values of the parameters.
In order to identify phase $+$ and phase $-$ for these situations, we need to look more into the phase structure of this model.

\subsection{Phase structure}

The dynamics of the phase
transition will depend on $\epsilon _{1}$ and $\epsilon _{2}$ only
through the relative value $\epsilon
_{2}-\epsilon _{1}$. In particular, the critical temperature, i.e., that
at which $p_{1}=p_{2}$, is given by the equation
\begin{equation}
c_{2}^{2}a_{2}T_{c}^{\nu _{2}}-c_{1}^{2}a_{1}T_{c}^{\nu _{1}}=\epsilon
_{2}-\epsilon _{1}.  \label{Tc}
\end{equation}
Without loss of generality, we shall consider $\nu_1\leq \nu_2$.

\begin{figure}[bth]
\centering
\epsfysize=5cm \leavevmode \epsfbox{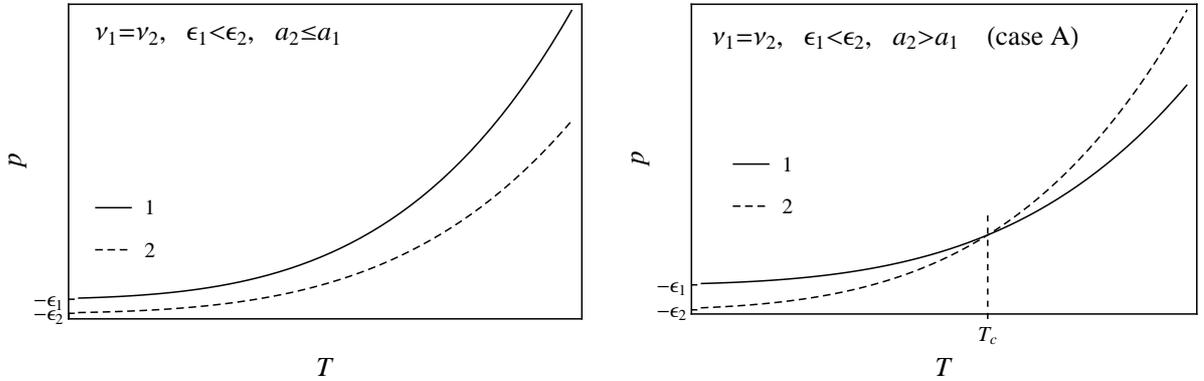}
\caption{The pressure as a function of
temperature for the two phases described by Eq. (\protect\ref{EOS}), with
$\protect\nu _{1}=\protect\nu _{2}$ and $\protect\epsilon _{1}<
\protect\epsilon _{2}$. For $a_{2}>a_{1}$ the model has a phase transition.}
\label{figpcc}
\end{figure}

\paragraph{Case A ($c_{+}=c_{-}$).}
Let us consider first the case $\nu _{1}=\nu _{2}$. In this case we may assume $\epsilon _{1}<\epsilon _{2}$ without loss of generality. Thus, for $a_{2}\leq a_{1}$ phase 1 is always the stable one (see Fig. \ref{figpcc}). Hence, there will be a phase transition only if $a_{2}>a_{1}$. For this phase transition, which we shall refer to as {Case A}, the situation is similar
to that with the bag EOS.
For $T>T_{c}$ we have $p_{2}(T)>p_{1}(T)$, whereas for $T<T_{c}$ we have $p_{2}(T)<p_{1}(T)$.
Hence, phase 2 dominates at high temperature and phase 1 dominates at low
temperature. We shall denote them with a ``$+$''
and a ``$-$'', respectively.
The speed of sound is the same in both phases, $c_{+}=c_{-}\equiv c_s$, only that now we may have $c_s\neq 1/\sqrt{3}$. The critical temperature is given by $T_{c}=\left[ \Delta \epsilon /(c_s^{2}\Delta a)\right] ^{1/\nu }$, with $\Delta a=a_{+}-a_{-},\Delta\epsilon=\epsilon_+-\epsilon_-$. The latent heat is given by
\begin{equation}
L=\nu \Delta \epsilon =\nu c_s^{2}\Delta aT_{c}^{\nu }.  \label{Lcc}
\end{equation}

\begin{figure}[hbt]
\centering
\epsfysize=5cm \leavevmode \epsfbox{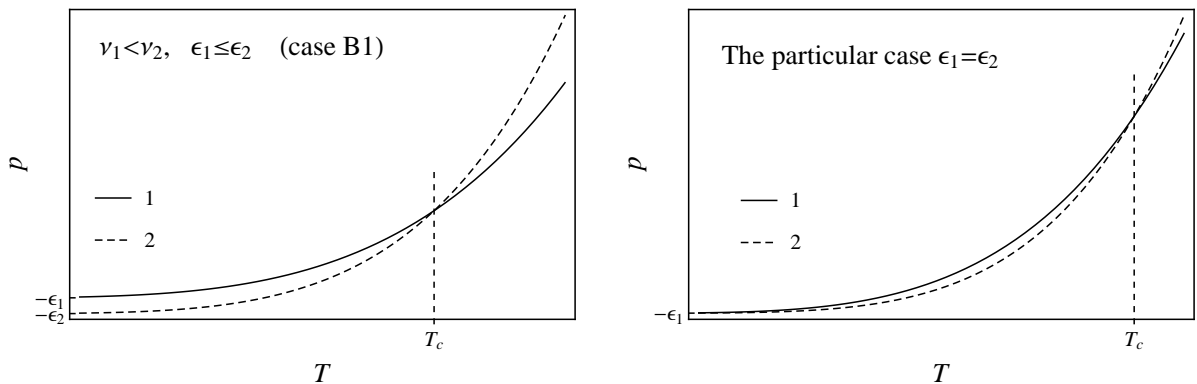}
\caption{The pressure as a function of
temperature for the two phases given by Eq. (\protect\ref{EOS}), with $
\nu _{1}<\protect\nu _{2}$ and $\protect\epsilon _{1}\leq
\epsilon _{2}$. There is a phase transition for any set of values of $
a_{+}$ and $a_{-}$.}
\label{figp1pt}
\end{figure}

\begin{figure}[hbt]
\centering
\epsfysize=5cm \leavevmode \epsfbox{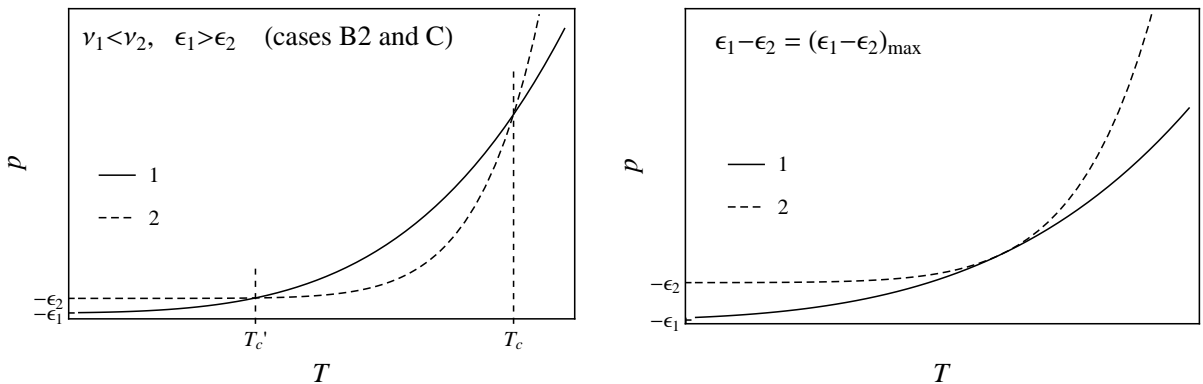}
\caption{The
pressure as a function of temperature for the two phases described by Eq.
(\ref{EOS}), with $\nu _{1}<\nu _{2}$ and $
\epsilon _{1}>\epsilon _{2}$. Under the conditions of Eq.
(\ref{cond2pt}), there are two phase
transitions.}
\label{figp2pt}
\end{figure}

\paragraph{Case B ($c_{+}<c_{-}$).}
If the exponents are different, say, $\nu _{1}<\nu _{2}$, we may have
several situations, depending on the relative values of $\epsilon_1$ and $\epsilon_2$. For $\epsilon _{1}\leq \epsilon _{2}$ (see Fig.
\ref{figp1pt}) there is a phase transition for any set of values of $a_{1}$
and $a_{2}$, which occurs at a temperature given by Eq. (\ref{Tc}). We shall
refer to this phase transition as \textbf{Case B1}.
We shall thus denote phase 2 (the high-temperature phase) with a ``$+$''
and phase 1 (the low-temperature phase) with a ``$-$''.
Therefore, we have $\nu _{-}<\nu _{+}$ and $c_{+}<c_{-}$. The latent
heat is given by
\begin{equation}
L=\frac{c_{-}^{2}-c_{+}^{2}}{c_{-}^{2}}a_{+}T_{c}^{\nu _{+}}+\nu _{-}\Delta
\epsilon ,  \label{latheat}
\end{equation}
with $\Delta \epsilon \equiv \epsilon _{+}-\epsilon
_{-}=\epsilon _{2}-\epsilon _{1}$.
We have a first-order phase transition (i.e, $L>0$) even in the case $\Delta \epsilon=0$ (see Fig. \ref{figp1pt}, right panel). In fact, if we consider negative values of $\Delta \epsilon$, we still have a first-order phase transition, since we have $L>0$ up to $\Delta \epsilon =-\frac{c_{-}^{2}-c_{+}^{2}}{\nu _{-}c_{-}^{2}}a_{+}T_{c}^{\nu _{+}}$ (see Fig.
\ref{figp2pt}, left panel).
This phase transition with $\Delta \epsilon <0$ still corresponds to case B ($c_{+}<c_{-}$), and we shall refer to it as \textbf{Case B2}.

\paragraph{Case C ($c_{-}<c_{+}$).}
As can be seen in Fig. \ref{figp2pt}, for $\Delta \epsilon <0$ we may have two phase transitions\footnote{In such a case, as the temperature of the Universe
decreases we have, first, a phase transition at $T=T_{c}$ and, then, a
second phase transition at $T=T_{c}^{\prime }$.}, i.e., Eq. (\ref{Tc}) may have two solutions $T_{c}$ and $T_{c}^{\prime}$. The phase transition at $T=T_c$ corresponds to case B2, and we shall refer to the phase transition at the smaller temperature $T_{c}^{\prime }$ as {Case C}.
In case C, the phase transition is from phase 1 back to
phase 2. Therefore, in this case we will denote phase 1 with a
``$+$'' and phase 2 with a ``$-$'', i.e., the roles of the phases are inverted with respect to the phase transition at $T_c$.
In particular, $\Delta \epsilon \equiv \epsilon
_{+}-\epsilon _{-}=\epsilon _{1}-\epsilon _{2}$ is now positive. The latent heat is given by
\begin{equation}
L^{\prime }=\frac{c_{-}^{2}-c_{+}^{2}}{c_{-}^{2}}a_{+}T_{c}^{\prime \nu
_{+}}+\nu _{-}\Delta \epsilon .  \label{latheatpri}
\end{equation}
This is the same expression as Eq. (\ref{latheat}),
but the critical temperature is now given by the other solution of Eq. (\ref{Tc}), and we have $\Delta \epsilon >0$,  $c_{-}^{2}-c_{+}^{2}<0$,
while in Eq. (\ref{latheat}) we had $\Delta \epsilon <0$, $
c_{-}^{2}-c_{+}^{2}>0$.  Case C is characterized by the relation $c_{-}<c_{+}$.
This phase transition  exists for $\nu _{1}<\nu _{2}$ and $
\epsilon _{1}>\epsilon _{2}$, provided that $\epsilon
_{1}-\epsilon _{2}$ is not too large, as can be appreciated in Fig. \ref{figp2pt}.
For $\epsilon _{1}-\epsilon _{2}=0$, we have $T_{c}^{\prime }=0$
(Fig. \ref{figp1pt}, right panel). As we increase $\epsilon
_{1}-\epsilon _{2}$, the two phase transitions approach each
other, ending in a single point (Fig. \ref{figp2pt}, right panel). This occurs
for\footnote{This expression can be obtained by demanding, besides the equation $p_{1}=p_{2}$ for the critical temperature, the condition $s_{1}=s_{2}$
(i.e., $dp_{1}/dT=dp_{2}/dT$).}
\begin{equation}
\left( \epsilon _{1}-\epsilon _{2}\right) _{\max }=\left[ \left( \frac{
\nu _{1}}{\nu _{2}}\right) ^{\nu _{1}}\frac{(c_{1}^{2}a_{1})^{\nu _{2}}}{
(c_{2}^{2}a_{2})^{\nu _{1}}}\right] ^{\frac{1}{\nu _{2}-\nu _{1}}}\frac{\nu
_{2}-\nu _{1}}{\nu _{2}}.  \label{cond2pt}
\end{equation}
For higher values of $\epsilon _{1}-\epsilon _{2}$, there is no phase
transition.

It is important to note that this simple EOS, just like the bag EOS, will not describe a physical model in the whole temperature range. For instance, in the case $\nu_1<\nu_2,\epsilon_1>\epsilon_2$, we have two phase transitions, namely, from phase 2 to phase 1, and then back to phase 2. Although two-step phase transitions are possible in Cosmology (see, e.g., \cite{kpsw14}), they will involve in general three different phases.
In spite of this, each of the cases above might provide
useful approximations for real phase transitions, at least in the small
temperature range in which the phase transition occurs.
In particular, from the discussion of Sec. \ref{EOSs} it seems likely that in a realistic case we will have $c_-<c_+$, which is described in our model by case C (the phase transition at $T=T'_c$). If extrapolated to higher temperatures, our model would give another phase transition at $T=T_c$ (case B2), but this will not occur in the physical model which is being approximated at $T=T_c'$.

\subsection{Hydrodynamics at the phase transition front}

Let us consider a moving bubble wall like in section \ref{hidrogral}, and
denote with subindexes ``$+$'' and
``$-$'' the values of the fluid velocity
just in front and just behind the wall discontinuity, respectively. The
interface conditions (\ref{land1}-\ref{land2}) give the relation (in the
reference frame of the wall)
\begin{equation}
v_{+}=\frac{q\left( \frac{v_{-}}{2}+\frac{c_{-}^{2}}{2v_{-}}\right) \pm
\sqrt{q^{2}\left( \frac{v_{-}}{2}+\frac{c_{-}^{2}}{2v_{-}}\right)
^{2}-\left( 1+\alpha _{+}\right) \left( c_{+}^{2}-\alpha _{+}\right) }}{
1+\alpha _{+}},  \label{vmavme}
\end{equation}
where
\begin{equation}
q=\frac{1+c_{+}^{2}}{1+c_{-}^{2}}
\end{equation}
and
\begin{equation}
\alpha _{+}=\frac{\Delta \epsilon }{a_{+}T_{+}^{\nu _{+}}},
\label{alfa+gral}
\end{equation}
with
\begin{equation}
\Delta \epsilon =\epsilon _{+}-\epsilon _{-}.
\end{equation}
This is the generalization of the well known relation for the bag EOS
\cite{s82}. Detonations correspond to the $+$ sign and deflagrations to the $-$
sign. Notice that for $\alpha _{+}>c_{+}^{2}$ we only have detonations. The
velocity $v_{+}$ has an extremum at $v_{-}=c_{-}$, given by $v_{+}=v_{J}$,
where
\begin{equation}
v_{J}^{\QATOP{\mathrm{det}}{\mathrm{def}}}(\alpha _{+})=\frac{qc_{-}\pm
\sqrt{q^{2}c_{-}^{2}-\left( 1+\alpha _{+}\right) \left( c_{+}^{2}-\alpha
_{+}\right) }}{1+\alpha _{+}}  \label{vj}
\end{equation}
is the Jouguet velocity. Notice that, if we regard $c_{\pm }$ as fixed parameters of
the model, the curves of $v_{+}$ vs $v_{-}$ will depend only on the
temperature-dependent variable $\alpha _{+}$, like in the bag case. The
relation between $w_{-}$ and $w_{+}$ is readily obtained from Eq. (\ref{land2}),
\begin{equation}
w_{-}/w_{+}=\left( v_{+}\gamma _{+}^{2}\right) /\left( v_{-}\gamma
_{-}^{2}\right) .  \label{wmewma}
\end{equation}
This ratio also depends only on $\alpha _{+}$.
We shall now analyze the relation between $v_{+}$ and $v_{-}$ for the three
kinds of phase transitions described above (namely, cases A, B, and C).

\subsubsection{Case A: $c_{-}=c_{+}$}

\begin{figure}[hbt]
\centering
\epsfysize=7cm \leavevmode \epsfbox{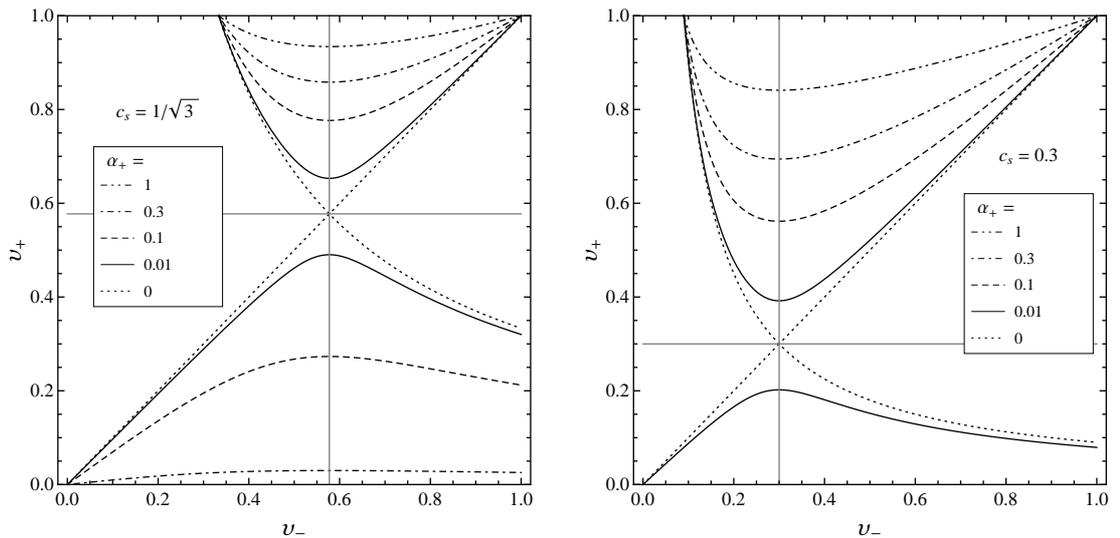}
\caption{$v_{+}$ vs $v_{-}$ for the case $c_{+}=c_{-}$. The horizontal and vertical gray lines indicate the values $v_\pm=c_\pm$.}
\label{figvmavmea}
\end{figure}

Case A is the simplest generalization of the bag EOS.
There is a single speed of sound, $c_{+}=c_{-}\equiv c_s$ (i.e., a single parameter $\nu
=\nu _{+}=\nu _{-}$), and we have $q=1$. We plot $v_{+}$ vs $v_{-}$ in Fig.
\ref{figvmavmea} for several values of $\alpha _{+}$. The left panel corresponds to the bag case $c_s=1/\sqrt{3}$. The general structure of the curves is similar for any value of $c_s$.
Upper curves ($v_{+}>c_s$) correspond to detonations and lower curves ($v_{+}<c_s$) correspond to deflagrations. Weak detonations correspond to $v_->c_s$ while weak deflagrations correspond to $v_-<c_s$.

In this simple case, the variable $\alpha _{+}$ is directly related to the
strength of the phase transition (since $\Delta \epsilon $ is
proportional to the energy density discontinuity $L$),
\begin{equation}
\alpha _{+}=\frac{L/\nu }{a_{+}T_{+}^{\nu }}=c_s^{2}\frac{\Delta a}{a_{+}}
\left( \frac{T_{c}}{T_{+}}\right) ^{\nu }.
\end{equation}
For a fixed $T_{+}$, higher values of $\alpha _{+}$ correspond to stronger
phase transitions, i.e., to higher values of $L$ or $\Delta a$, whereas for
weakly first-order phase transitions (small $L$ and $\Delta a$) we will have
a small $\alpha _{+}$. Besides, $\alpha _{+}$ increases as $T_{+}$
decreases, i.e., as the amount of supercooling increases. Therefore, we
expect strong departures from equilibrium for high values of $\alpha _{+}$
and smaller departures for small $\alpha _{+}$. This is reflected in Fig.
\ref{figvmavmea}. The higher the value of $\alpha _{+}$, the higher the
difference between $v_{+}$ and $v_{-}$. This means that, as expected,
perturbations caused by the wall on the fluid are stronger for stronger
phase transitions. On the contrary, for $\alpha _{+}\rightarrow 0$, the
curves (in the weak regions) approach the line of $v_{+}=v_{-}$.

Although a small $T_{+}$ implies a strong supercooling, the converse is not
true, as there may be reheating in front of the wall (i.e., $T_+>T_n$). Thus, in some cases we
may have $T_{+}\simeq T_{c}$ and even $T_{+}>T_{c}$. In any case, $T_{+}$
will never be much higher than $T_{c}$. Notice that the exact point $
T_{+}=T_{c}$, which corresponds to the value $\alpha _{+}=c_s^{2}\Delta
a/a_{+}\equiv \alpha _{c}$, is not a special case for hydrodynamics according to Eqs.
(\ref{vmavme}-\ref{vj}). For $T_{+}\ll T_{c}$ (strong supercooling) we may
have $\alpha _{+}\gg 1$, while for $T_{+}\approx T_{c}$ (small or moderate
supercooling) the value of $\alpha _{+}$ will depend essentially on the
parameter $L/w_{+}$.
In the limit $\alpha _{+}=0$ the detonation curve and the deflagration curve
touch each other at the Jouguet point. In principle, this limit would
correspond to a second-order phase transition (since $L=\Delta a=0$).
However, if we consider a fixed $T_{c}$ and take this limit,
then we are left with a single phase (see Fig. \ref{figpcc}).
Hence, in this limit the model gives no phase transition at all.

\subsubsection{Case B: $c_{-}>c_{+}$}

\begin{figure}[hbt]
\centering
\epsfysize=7cm \leavevmode \epsfbox{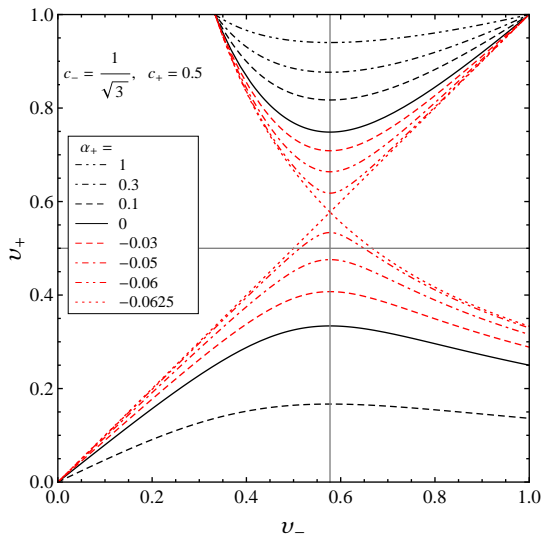}
\caption{$v_{+}$ vs $v_{-}$ for $
c_{-}>c_{+}$. The horizontal and vertical grey lines indicate the values  $v_{\pm}=c_{\pm}$.}
\label{figvmavmeb}
\end{figure}

A peculiarity of case B is that we may have $\Delta \epsilon <0$ and,
thus, $\alpha _{+}<0$. Nevertheless, according to  Eq. (\ref{latheat}), in this case $\Delta\epsilon$ is no longer
proportional to the latent heat. Therefore, it is not directly related to the
strength of the phase transition. As a consequence, in the limit $\alpha
_{+}=0$ we will still have a first-order phase transition with $L>0$. This  can be
seen also in Fig. \ref{figp1pt} (right panel).
We may write
\begin{equation}
\alpha _{+}=\frac{L/\nu _{-}}{a_{+}T_{+}^{\nu _{+}}}-\frac{
c_{-}^{2}-c_{+}^{2}}{1+c_{-}^{2}}\left( \frac{T_{c}}{T_{+}}\right) ^{\nu
_{+}}.  \label{alfa+b}
\end{equation}
The curves of $v_{+}$ vs $v_{-}$ for this case are plotted in Fig.
\ref{figvmavmeb}. Black curves correspond to positive values of $\alpha _{+}$
(case B1), while red curves correspond to negative values of $\alpha _{+}$
(case B2).

As in case A, different curves may correspond either to different values of
the parameters or to different values of the variable $T_{+}$, due to the
dependence on a single parameter combination (namely, the variable $\alpha
_{+}$). Case B1 ($\Delta \epsilon >0$) behaves pretty much like case A.
Namely, high values of $\alpha _{+}$ correspond either to large values of $L$
or large amounts of supercooling. As a consequence, a large $\alpha_+$ gives a large difference between $v_{+}$ and $v_{-}$, and the strength of hydrodynamics decreases as $\alpha_+$ decreases. However, the limit $\alpha _{+}=0$ still corresponds to a finite $L$ and nothing relevant happens to the hydrodynamics\footnote{For $\alpha_+=0$, the relations $v_{+}$-$v_{-}$ and $w_{+}$-$w_{-}$ do not depend on the value of $T_+$. However, other discontinuities will not be given by the single combination of parameters $\alpha_+$. For instance, the relation between $T_{-}$ and $T_{+}$, must be obtained from $w_{\pm }=(1+c_{\pm }^{2})a_{\pm }T_{\pm }^{\nu _{\pm }}$.}.
In this limit, case B1 matches case B2.
In case B2 ($\Delta \epsilon <0$) the strength of the hydrodynamics
continues decreasing for decreasing $\alpha _{+}$, which is now negative.
The limit of an extremely weak hydrodynamics (corresponding to the
limiting curve $v_{+}=v_{-}$), is achieved for $\alpha _{+}=\alpha _{w}$, where
\begin{equation}
\alpha _{w}=-(c_{-}^{2}-c_{+}^{2})/(1+c_{-}^{2}).  \label{alfaweak}
\end{equation}

Although $\epsilon_\pm $ is the energy density of
false vacuum, the interpretation of $\Delta\epsilon$ as the vacuum energy that is released in the phase transition is far from clear, as well as that of $a_{+}T_{+}^{\nu _{+}}$ as  radiation or thermal
energy. Thus, a more physical variable, rather than $\alpha _{+}$,
would be the ratio of physical quantities $L/w_{+}$. Eq. (\ref{alfa+b})
shows that $\alpha _{+}$ depends separately on $L/w_{+}$ and the amount of
supercooling $T_{c}/T_{+}$ (in contrast, for case A we have $\alpha
_{+}\propto L/w_{+}$). We define the parameter
\begin{equation}
\bar{L}\equiv \frac{L}{w_{+}(T_{c})}=\frac{L}{(1+c_{+}^{2})a_{+}T_{c}^{\nu
_{+}}}.  \label{lbar}
\end{equation}
In terms of physical quantities, the weak limit $\alpha _{+}=\alpha _{w}$ is
obtained for
\begin{equation}
1-\left[ \frac{T_{w}}{T_{c}}\right] ^{\nu _{+}}=\frac{c_{-}^{2}(1+c_{+}^{2})
}{c_{-}^{2}-c_{+}^{2}}\bar{L},
\end{equation}
which implies $T_{w}<T_{c}$.
This means that the dotted curves in Fig. \ref{figvmavmeb} will be obtained not only in the limit $L=0$, but also with $L>0$ for a certain amount of supercooling.
 Thus,
for case B2 the strength of hydrodynamics decreases as the
supercooling increases. This is because the  second critical temperature $T'_c$ is approached.
 As a consequence, the hydrodynamics may become rather
strange (in comparison to the more familiar behavior of case A) near the
limiting value $\alpha _{+}=\alpha _{w}$. Notice, in particular, that for
deflagrations we may have $v_{+}>c_{+}$, which never occurs in case A.
Although it would be interesting to study the hydrodynamics for $T_{+}$ close to $T_{w}$, we will now argue that it is unlikely that a physical system would actually reach such a situation.

\begin{figure}[hbt]
\centering
\epsfysize=5cm \leavevmode \epsfbox{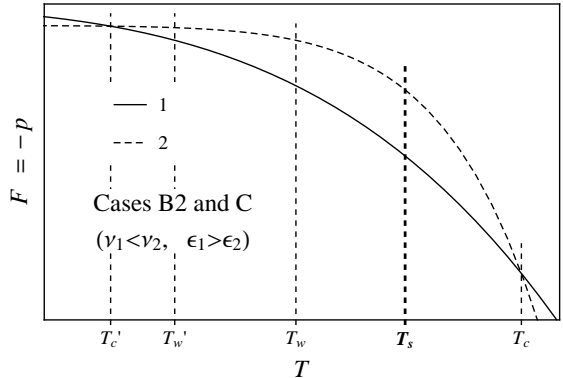}
\caption{The free energy density $\mathcal{F}(T)=-p(T)$ corresponding to the left panel of
Fig. \protect\ref{figp2pt}.}
\label{figtemps}
\end{figure}

Supercooling occurs because there is a barrier
between the minima $\phi _{\pm }$ of the free energy $\mathcal{F}(\phi ,T)$.
It is important to remark that our phenomenological model only describes the
thermodynamical quantities \emph{at the minima}
and does not have information on the barrier  separating them.
Nevertheless, we may guess some information on the possible amount of supercooling from the
separation of the values  $\mathcal{F}_{\pm }(T)=\mathcal{F}\left( \phi _{\pm }(T),T\right) $. At
the critical temperature $T_{c}$, the two minima are degenerate, i.e., $
\mathcal{F}_{-}(T_{c})=\mathcal{F}_{+}(T_{c})$, and nucleation is
impossible. Below $T_{c}$, $\phi _{-}$ becomes the absolute minimum of $
\mathcal{F}$, and phase $+$ becomes metastable. Thus, we have $\mathcal{F}
_{-}(T)<\mathcal{F}_{+}(T)$.
In general, as $
T $ descends and the value $\mathcal{F}_{-}(T)$ moves away from $\mathcal{F}
_{+}(T)$, the barrier between the minima gets shorter and nucleation becomes
more probable. Bubble nucleation will effectively begin when a certain
amount of supercooling is reached, such that the barrier is small enough and
the values $\mathcal{F}_{\pm }(T)$ are  separated enough. In some cases it
may happen that the barrier never gets small enough and the system remains
stuck in the metastable phase. In our case B2, the barrier cannot become
arbitrarily small, since there is another first-order phase transition at
temperature $T_{c}^{\prime }$. Below a certain temperature $T_{s}$, $
\mathcal{F}_{-}(T)$ and $\mathcal{F}_{+}(T)$ approach each other again,
since at $T=T_{c}^{\prime }$ the free energy is again degenerate (see Fig.
\ref{figtemps}). As this happens, the nucleation becomes less probable,
since at $T=T_{c}^{\prime }$ it is not possible at all. Hence, we expect
that, if bubble nucleation does not begin by the time $T$ decreases below $
T_{s}$, then the phase transition will never happen.

Thus, the temperature $T_{s}$
is a physical bound for the nucleation temperature $T_{n}$.
The  maximum separation of the values $
\mathcal{F}_{\pm }(T)$ is given by the condition $s_{1}(T_{s})=s_{2}(T_{s})$
. This gives
\begin{equation}
T_{s}^{\nu _{2}-\nu _{1}}=\frac{1+c_{1}^{2}}{1+c_{2}^{2}}\frac{a_{1}}{a_{2}}.
\end{equation}
It is not difficult to show that we will always have $T_{w}<T_{s}$, as
indicated in Fig. \ref{figtemps}. Defining $\alpha _{s}=\Delta \epsilon
/(a_{+}T_{s}^{\nu _{+}})$ and using the condition (\ref{cond2pt}) for $
|\Delta \epsilon |$ and Eq. (\ref{alfaweak}) for $\alpha _{w}$, we
readily obtain $|\alpha _{s}|<|\alpha _{w}|$. This implies that $T_{s}>T_{w}$.
Moreover, the weak-hydrodynamics limit $T_{s}=T_{w}$ is only reached at
the maximum of $|\Delta \epsilon |$, i.e., in the limit $L=0$, in which
the phase transitions disappear (Fig. \ref{figp2pt}, right panel). We will
assume that the nucleation temperature fulfills $T_{n}\geq T_{s}$ and,
hence, we will not worry about getting close to the weak limit in case B.

\subsubsection{Case C: $c_{-}<c_{+}$}

\begin{figure}[hbt]
\centering
\epsfysize=7cm \leavevmode \epsfbox{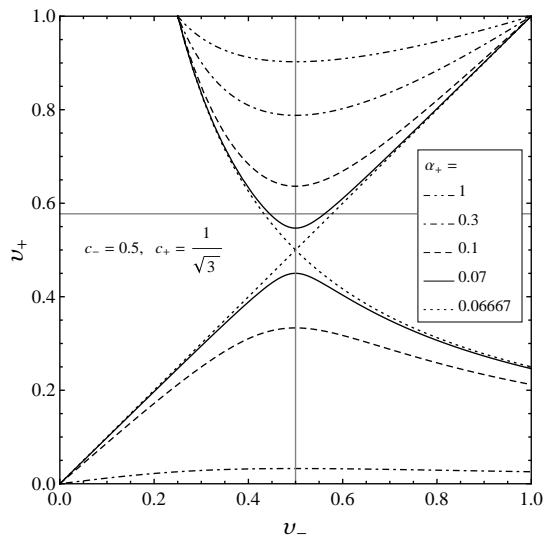}
\caption{$v_{+}$ vs $v_{-}$ for $
c_{-}<c_{+}$. The horizontal and vertical grey lines indicate the values  $v_{\pm}=c_{\pm}$.}
\label{figvmavmec}
\end{figure}

For the phase transition at temperature $T_{c}^{\prime }$ we have $
c_{-}<c_{+}$, with $c_{-}=c_{2}$ and $c_{+}=c_{1}$. In this case we have $
\Delta \epsilon >0$ (see Fig. \ref{figp2pt}) and, hence, the variable $\alpha
_{+}\equiv \Delta \epsilon /(a_{+}T_{+}^{\nu _{+}})$ is always
positive. Nevertheless, the weak-hydrodynamics limit is
reached in this case for a positive value\footnote{Although Eqs.
(\ref{alfaweak}) and (\ref{alfaweakpri}) seem the same, the
subindexes $\pm $ have different meanings in each case. We have $\alpha
_{w}=-(c_{1}^{2}-c_{2}^{2})/(1+c_{1}^{2})$ and $\alpha _{w}^{\prime
}=(c_{1}^{2}-c_{2}^{2})/(1+c_{2}^{2})$. Notice that the numerators are
opposite and the denominators are just different.},
\begin{equation}
\alpha _{w}^{\prime }=(c_{+}^{2}-c_{-}^{2})/(1+c_{-}^{2}).
\label{alfaweakpri}
\end{equation}
We may  write the variable $\alpha_+$ as
\begin{equation}
\alpha _{+}=\frac{L^{\prime }/\nu _{-}}{a_{+}T_{+}^{\nu _{+}}}+
\frac{c_{+}^{2}-c_{-}^{2}}{1+c_{-}^{2}}\left( \frac{T_{c}^{\prime }}{T_{+}}
\right) ^{\nu _{+}}.
\end{equation}
The curves of $v_{+}$ vs $v_{-}$ for this case are plotted in Fig. \ref{figvmavmec}.

Like in case B2, we see that the hydrodynamics may become unfamiliar near
the weak limit. As can be seen in Figs. \ref{figvmavmea}, \ref{figvmavmeb} and
\ref{figvmavmec}, in the detonation and deflagration curves the value of  $v_{+}$ is bounded by $c_{-}$ rather than by $c_{+}$. This is because this bound is
given by the conditions $v_{+}=v_{J}(\alpha _{+})$ (the extremum of $v_{+}$)
and $v_{+}=v_{-}$ (the weak limit). The former condition corresponds to $
v_{-}=c_{-}$ and, hence, the latter gives $v_{+}=c_{-}$. In the present
case, since the speed of sound is higher in the $+$ phase, we may have
detonations with $v_{+}<c_{+}$, as can be observed in Fig. \ref{figvmavmec}.
This would correspond to detonations which are subsonic with respect to the
fluid in the $+$ phase.
Such subsonic detonations are in principle possible near the weak limit. It
can be seen that this limit corresponds to a temperature $T_{w}^{\prime}>T_{c}'$, as indicated  in Fig. \ref{figtemps}. Indeed, writing Eq.
(\ref{alfaweakpri}) in terms of physical quantities, we have
\begin{equation}
\left[ \frac{T_{w}^{\prime }}{T_{c}^{\prime }}\right] ^{\nu _{+}}-1=\frac{
c_{-}^{2}(1+c_{+}^{2})}{c_{-}^{2}-c_{+}^{2}}\bar{L}^{\prime }>0.
\end{equation}
These detonations, for which the incoming flow is subsonic, will
be preceded by a shock wave which reheats the fluid in front of the wall.
Hence, we do not have the familiar restriction $T_{+}=T_{n}<T_{c}$.

\subsection{The shock discontinuity}

Let us now consider a shock front. The discontinuities of the fluid
variables at this front are given by Eqs. (\ref{land1}-\ref{land2}). We
shall use the index 1 for fluid variables behind the shock and the index 2
for variables in front of the shock. Thus, in the shock frame we have an
incoming velocity $-v_{2}$ and an outgoing velocity $-v_{1}$. Eqs.
(\ref{land1}-\ref{land2}) can be written in the form \cite{landau}
\begin{equation}
v_{1}v_{2}=\frac{p_{2}-p_{1}}{e_{2}-e_{1}},\quad \frac{v_{1}}{v_{2}}=\frac{
e_{2}+p_{1}}{e_{1}+p_{2}}.
\end{equation}
Since the EOS is the same on both sides of the discontinuity, these give
\begin{equation}
v_{1}v_{2}=c_s^{2},\quad \frac{v_{1}}{v_{2}}=\frac{T_{2}^{\nu} +
c_s^{2}T_{1}^{\nu }}{T_{1}^{\nu }+c_s^{2}T_{2}^{\nu }}.  \label{landaushock}
\end{equation}
Notice that the first of Eqs. (\ref{landaushock}) implies that, in the frame
of the shock front, one of the fluid velocities is subsonic and the other
supersonic.

It is possible to obtain important constraints on the fluid velocities by
requiring the entropy of a fluid element to increase as it passes through a
discontinuity surface (see, e.g., \cite{gkkm84}). In particular, for the
shock discontinuity it can be shown that $v_{1}<c_s<v_{2}$ (see, e.g.,
\cite{lm11}). In the frame of the bubble center, the fluid velocities on each
side of the shock front are given by $\tilde{v}_{1,2}=(v_{\mathrm{sh}}- v_{1,2})/(1-v_{\mathrm{sh}}v_{1,2})$. Hence, the condition $v_{1}<v_{2}$ implies $\tilde{v}_{1}>\tilde{v}_{2}$, i.e., the fluid velocity must have a
negative jump. Due to the boundary condition of vanishing fluid velocity far
behind the wall (i.e., at the bubble center), such a negative jump can only
occur in front of the wall (see Ref. \cite{lm11} for a detailed discussion).
Therefore, \emph{a shock front can only propagate in front of the phase
transition front}.

On the other hand, in front of the shock we have $\tilde{v
}_{2}=0$ (unless there is another source of fluid motions, such as the wall
of another bubble). Hence, the shock front propagates with a velocity given
by $v_{\mathrm{sh}}=v_{2}$. Therefore, the first of Eqs. (\ref{landaushock})
gives the relation
\begin{equation}
\tilde{v}_{1}=\frac{c_+^{2}}{1-c_+^{2}}\frac{v_{\mathrm{sh}}^{2}/c_+^{2}-1}{v_{\mathrm{sh}}},
\label{v1t}
\end{equation}
where we have taken into account the fact that the shock propagates in the $+$ phase.
Equivalently, we have
\begin{equation}
v_{\mathrm{sh}}=\frac{1-c_+^{2}}{2}\tilde{v}_{1}+\sqrt{\left( \frac{1-c_+^{2}}{2}
\tilde{v}_{1}\right) ^{2}+c_+^{2}}.  \label{xish}
\end{equation}
These equations show that the shock front is supersonic, $v_{\mathrm{sh}}>c_+$.
On the other hand, the second of Eqs. (\ref{landaushock}) gives the
relation
\begin{equation}
\frac{T_{1}^{\nu_+}-T_{2}^{\nu_+}}{\sqrt{T_{1}^{\nu_+}+c_+^{2}T_{2}^{\nu_+}} \sqrt{T_{2}^{\nu_+}+c_+^{2}T_{1}^{\nu_+}}}=\frac{\tilde{v}_{1}}{c_+}.  \label{relt1t2}
\end{equation}
This shows that the fluid is reheated behind the shock front, i.e., $
T_{1}>T_{2}$. Equivalently, we have
\begin{equation}
\left( \frac{T_{2}}{T_{1}}\right)^{\nu_+}=
\frac{c_+^{2}(1-v_{\mathrm{sh}}^{2})}{v_{\mathrm{sh}}^{2}-c_+^{4}}.  \label{relt1t2b}
\end{equation}

\subsection{Fluid profiles}

The solutions of the fluid equations are quite simple if the speed of sound is a constant. In
order to avoid confusion, we shall denote with a tilde fluid velocities in
the reference frame of the bubble center. Thus, according to Eqs.
(\ref{vconst}-\ref{vrar}), we have either constant solutions or the rarefaction solution
\begin{equation}
\tilde{v}_{\mathrm{rar}}(\xi )=\frac{\xi -c_{-}}{1-c_{-}\xi }.  \label{vrar2}
\end{equation}
We have set $c_s=c_{-}$ for the rarefaction, since this solution is physically
possible only behind the wall (assuming the wall is the only source of fluid motions). For constant $\tilde v$ we have constant enthalpy density, while for the rarefaction $w$ is given by Eq. (\ref{enth}),
\begin{equation}
w_{\mathrm{rar}}=w_{0}\exp \left[ \nu _{-}c_{-}\int_{\tilde{v}_{0}}^{\tilde{v
}_{\mathrm{rar}}}\frac{d\tilde{v}}{1-\tilde{v}^{2}}\right] =w_{0}\left(
\frac{1-\tilde{v}_{0}}{1+\tilde{v}_{0}}\frac{1+\tilde{v}_{\mathrm{rar}}}{1-
\tilde{v}_{\mathrm{rar}}}\right) ^{\frac{c_{-}\nu _{-}}{2}}, \label{enthanal}
\end{equation}
where the values $w_{0}=w(\xi _{0})$ and $\tilde{v}_{0}=\tilde{v}(\xi _{0})$
must be chosen according to the boundary and matching conditions. Inserting Eq. (\ref{vrar2}) in Eq. (\ref{enthanal}), we obtain
\begin{equation}
\frac{w_{\mathrm{rar}}}{w_{0}}=\left( \frac{1-c_{-}}{1+c_{-}}\,\frac{1-
\tilde{v}_{0}}{1+\tilde{v}_{0}}\,\frac{1+\xi }{1-\xi }\right) ^{c_{-}\nu
_{-}/2}.  \label{enthprof}
\end{equation}

The boundary conditions are that the fluid velocity vanishes far behind and
far in front of the wall, and that the temperature far in front of the wall
is given by the nucleation temperature $T_{n}$ [therefore, the enthalpy density is given by $w_n=w_+(T_n)$]. On the other hand, we have
matching conditions for the wall and shock fronts.

The values of the fluid velocity just in front and just behind the phase
transition discontinuity at $\xi _{w}=v_{w}$ are given by
\begin{equation}
\tilde{v}_{\pm }=\frac{\xi _{w}-v_{\pm }}{1-\xi _{w}v_{\pm }},
\label{vmametil}
\end{equation}
where $v_{\pm }$ are related by Eq. (\ref{vmavme}). We may also have a shock
front at a position $\xi _{\mathrm{sh}}=v_{\mathrm{sh}}>\xi _{w}$, i.e., in
the $+$ phase. In front of the shock the fluid velocity vanishes, $\tilde{v}
_{2}=0$, while behind the shock the fluid velocity is given by Eq. (\ref{v1t}).
The shock velocity $v_{\mathrm{sh}}$ and the fluid velocity $\tilde{v}_{1}$
can be obtained as functions of the temperature using Eq. (\ref{relt1t2b}).
In front of the shock we generally have $T_{2}=T_{n}$, and behind it $
T_{1}=T_{+}$ (see below). Therefore, we have the relation
\begin{equation}
\frac{c_{+}^{2}(1-v_{\mathrm{sh}}^{2})}{v_{\mathrm{sh}}^{2}-c_{+}^{4}}=\frac{w_n}{w_+}=\frac{\alpha _{+}}{\alpha _{n}},
\label{a1an}
\end{equation}
with
\begin{equation}
\alpha _{n}\equiv \Delta \epsilon /(a_{+}T_{n}^{\nu _{+}}). \label{an}
\end{equation}

\begin{figure}[hbt]
\centering
\epsfysize=5cm \leavevmode \epsfbox{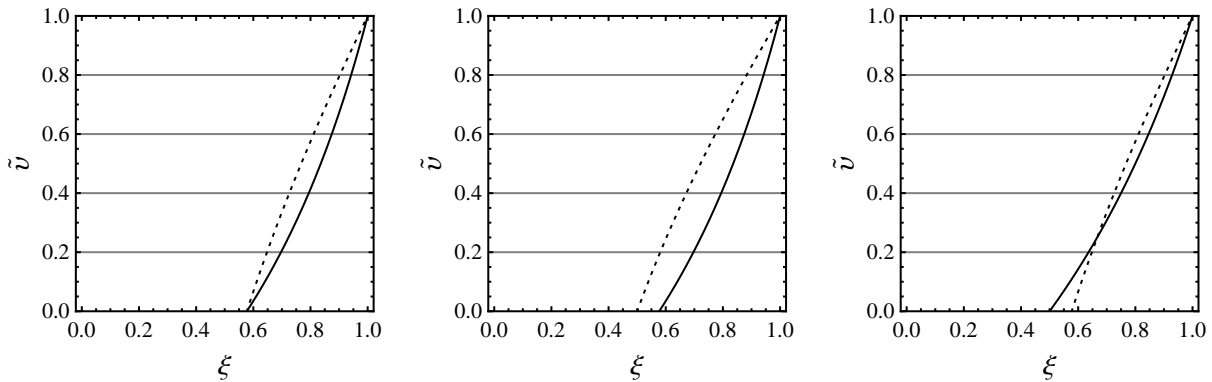}
\caption{The solution $\tilde{v}_{\mathrm{rar}}(\xi )$ (solid lines)
and the curve of points $(\xi_{\mathrm{sh}},\tilde{v}_{1})$ (dotted
lines), for $c_{+}=c_{-}=1/\sqrt{3}$ (left panel),
$c_{+}=0.5,c_{-}=1/\sqrt{3}$ (central panel), and
$c_{+}=1/\sqrt{3}, c_{-}=0.5$ (right panel).
We have indicated also some of the constant solutions.}
\label{figsols}
\end{figure}

We will now consider the different kinds of profiles which can be
constructed using these solutions and conditions. To help the construction, it is useful to plot the solutions for $\tilde v(\xi)$ together with the curve of points $(\xi_{\mathrm{sh}},\tilde{v}_{1})$. The latter is given by Eq. (\ref{v1t}) with $v_{\mathrm{sh}}=\xi_{\mathrm{sh}}$. In Fig. \ref{figsols} we consider this plot for the three types of phase transitions
discussed above, namely, $c_{+}=c_{-}$ (case A), $c_{+}<c_{-}$ (case B), and $c_{+}>c_{-}$ (case C).

\subsubsection{The traditional detonation}

The boundary condition of a vanishing fluid
velocity far in front of the wall can only be achieved through a
discontinuity. This discontinuity may be either the phase transition front
or a shock front. In the former case, we have $\tilde{v}_{+}=0$, i.e., the
fluid is unperturbed in front of the wall.
Let us first consider this case. Therefore, we have $v_{w}=v_{+}$, and behind the wall
we must have a velocity $\tilde{v}_{-}>0$. Thus, we have $\tilde{v}_{-}>
\tilde{v}_{+}$ and, hence, $v_{+}>v_{-}$. Therefore, the hydrodynamical
process is a \emph{detonation}.

\begin{figure}[hbt]
\centering
\epsfysize=4cm \leavevmode \epsfbox{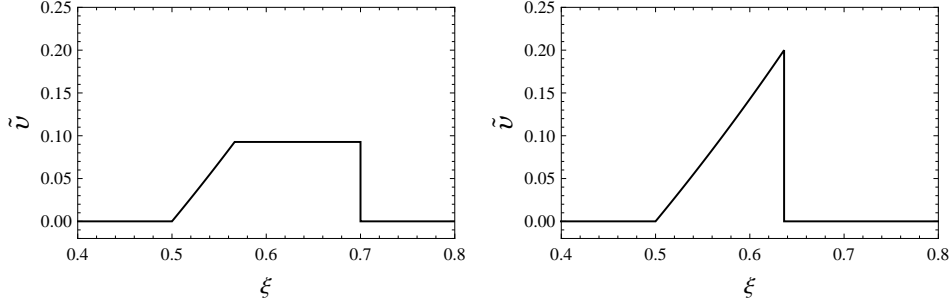}
\caption{The fluid velocity profile for a weak
detonation wall with velocity $v_{w}=0.7>v_{J}^{\det }(\alpha _{n})$ (left
panel) and a Jouguet detonation with velocity $v_{w}=v_{J}^{\det }
(\alpha _{n})\simeq 0.64$ (right panel), for the case $c_{+}=1/\sqrt{3},c_-=0.5$,
and for $\alpha _{n}=0.1$.}
\label{figwdeto}
\end{figure}

As we have seen, for a detonation we always have $v_{+}>c_{-}$. As a
consequence, the wall is supersonic with respect to the bubble center,
$v_{w}>c_{-}$. Behind the wall, the fluid velocity must decrease
from $\tilde{v}_{-}$ to $0$ in order to fulfill the boundary condition at
the bubble center. Since we have $\xi _{w}>c_{-}$, this can be accomplished
by using the rarefaction solution (see Fig. \ref{figsols}). We show this construction in Fig. \ref{figwdeto}. The rarefaction solution matches the value $\tilde{v}=0$ at $\xi=c_{-}$ and the value $\tilde{v}_{-}$ at a point $\xi _{0}$ given by
\begin{equation}
\xi _{0}=\frac{\tilde{v}_{-}+c_{-}}{1+\tilde{v}_{-}c_{-}}.
\end{equation}
This profile requires the condition $\tilde{v}_{-}\leq v_{\mathrm{rar}
}\left( \xi _{w}\right) $. The equality corresponds to a limiting profile like the one shown in the right panel of Fig. \ref{figwdeto}. This condition is equivalent to $v_{-}\geq c_{-}$. This
means that the detonation is either a \emph{weak detonation}, with $v_->c_-$ and $v_{w}>v_{J}^{\det }(\alpha _{+})$, or a \emph{Jouguet detonation} with
velocity $v_{w}=v_{J}^{\det }(\alpha _{+})$. A strong detonation cannot exist since its profile cannot be
formed. For the Jouguet detonation, we have $\xi _{0}=\xi _{w}$.

The plots shown in Fig. \ref{figwdeto} correspond
to case C. For the other two cases the shapes of the profiles are similar
(including those of case B with negative values of $\alpha _{n}$).
The value of $\tilde{v}_{-}$ can be obtained as a function of the wall
velocity and the nucleation temperature as follows. Since in this case the
fluid is unperturbed in front of the wall, we have $\alpha _{+}=\alpha _{n}$. Therefore, the value of $v_{-}$ can be obtained as a
function of $v_{w}$ and $\alpha _{n}$ by inverting Eq. (\ref{vmavme}) (see
appendix \ref{flprof}). Then, $\tilde{v}_{-}$ is given by $\tilde{v}
_{-}=(v_{w}-v_{-})/(1-v_{w}v_{-})$. The enthalpy profile between $c_{-}$ and
$\xi _{0}$ is given by Eq. (\ref{enthprof}), with boundary conditions $
\tilde{v}_{0}=\tilde{v}_{-}$ and $w_{0}=w_{-}$ at $\xi =\xi _{0}$. The value
of $w_{-}$ is related to $w_{+}=w_n$ through Eq. (\ref{wmewma}),
\begin{equation}
w_{-}=\frac{v_{w}\gamma _{w}^{2}}{v_{-}\gamma _{-}^{2}}w_{n}.  \label{wme}
\end{equation}
From Eqs. (\ref{EOSgral}) and (\ref{an}), we have
\begin{equation}
w_{n}=\left( 1+c_{+}^{2}\right) a_{+}T_{n}^{\nu _{+}}.  \label{wn}
\end{equation}
For concrete applications, though, it may be more useful to leave the results in terms of $w_n$.

\subsubsection{The traditional deflagration}

If the phase transition front is subsonic with respect to the bubble
center, $\xi _{w}<c_{-}$, we must have $\tilde{v}_{-}=0$, since the
solution $v_{\mathrm{rar}}$ cannot be used to match the boundary condition
at $\xi =0$, and neither can exist a shock discontinuity with a positive
velocity jump. In this case we have $\tilde{v}_{+}>0$, which
implies $v_{+}<v_{-}$, and the process is a \emph{deflagration}.
Furthermore, the condition $\tilde{v}_{-}=0$ implies $v_{-}=v_{w}<c_{-}$,
and we have a \emph{weak deflagration}. In the limiting case $v_{w}=c_{-}$,
we have a Jouguet deflagration.

\begin{figure}[hbt]
\centering
\epsfysize=4cm \leavevmode \epsfbox{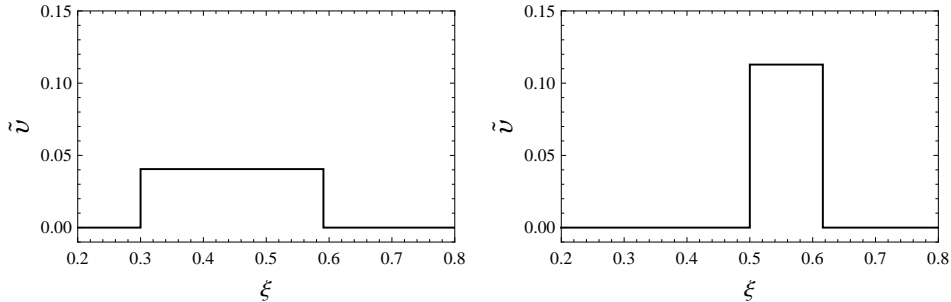}
\caption{The fluid velocity profiles for a
weak deflagration front with a velocity $v_{w}=0.3<c_{-}$ (left panel) and for a Jouguet
deflagration with velocity $v_{w}=c_{-}$ (right panel), for the case $c_{+}=1/\sqrt{3}, c_-=0.5$, and for $\alpha _{n}=0.1$.}
\label{figwdefla}
\end{figure}

A  look at Fig. \ref{figsols} shows that, in front of the wall, the
only possible solution is a constant $\tilde{v}=\tilde{v}_{+}$, and the
profile must end in a shock front. This profile is shown in Fig.
\ref{figwdefla}. The plots correspond to case C. The shapes of the profiles are
similar for the other two cases. For case B the profiles tend to be thinner and
taller, as can be expected from Fig. \ref{figsols}. Physically, this happens
because the shock front in that case propagates at a relatively lower velocity. As can be seen in the figure, for weak deflagrations the shock velocity is close to the speed of sound, $v_{\mathrm{sh}}\simeq c_+$.

The value of $\tilde{v}_{+}$ and the shock position $\xi _{\mathrm{sh}}$ can
be obtained as functions of $v_{w}$ and $\alpha _{n}$ as follows. In the
first place, $\xi _{\mathrm{sh}}$ has a one to one relation with $\tilde{v}
_{1}$ from Eqs. (\ref{v1t}-\ref{xish}), and the latter is given by $\tilde{v}
_{1}=$ $\tilde{v}_{+}$. It is easy to obtain $\tilde{v}_{+}$ as a function
of $v_{w}$ and $\alpha _{+}$, since we have $\tilde{v}_{+}=(v_{w}-$ $
v_{+})/(1-v_{w}v_{+})$, and $v_{+}$ is given by Eq. (\ref{vmavme}) as a
function of $\alpha _{+}$ and $v_{-}=v_{w}$. We only need to determine the
amount of reheating, i.e., the value of $\alpha _{+}$ as a function of $
\alpha _{n}$. This is given by Eq. (\ref{a1an}). Hence, we can eliminate the
variable $\alpha _{+}$ and obtain an equation for $\tilde{v}_{+}$ as a function of $\alpha _{n}$ and $v_{w}$ (see appendix \ref{flprof}).
The enthalpy in the shock-wave region is given by $w_{+}/w_{n}=\alpha
_{n}/\alpha _{+}$. On the other hand, the enthalpy behind the wall is given
by $w_{-}/w_{+}=(v_{+}\gamma _{+}^{2})/(v_{w}\gamma _{w}^{2})$.

This profile can also be constructed for a {supersonic wall}. However, in
this case the condition $\tilde{v}_{-}=0$ implies a strong deflagration ($v_{-}>c_{-}$), which is known to be unstable \cite{hkllm,mm14}.

\subsubsection{The supersonic deflagration}

The above kinds of solutions correspond to wall velocities in the ranges $
v_{w}>v_{J}^{\det }(\alpha _{n})$ or $v_{w}<c_{-}$. In general, there is a
gap between these two values. In Ref. \cite{kl95} it was shown that
supersonic deflagrations exist. The hydrodynamic solution consists of a Jouguet deflagration, and the wall is preceded by a shock front and is followed by a
rarefaction wave. Thus, the wall moves at the speed of sound with
respect to the fluid behind it,  $v_{-}=c_{-}$. However, since the
fluid has a velocity $\tilde{v}_{-}$ with respect to the bubble center, the
wall velocity is given by $v_{w}=(c_{-}+\tilde{v}_{-})/(1+c_{-}\tilde{v}
_{-})>c_{-}$. For the bag EOS, this kind of solutions fill the velocity gap
between $c_{-}$ and $v_{J}^{\det }(\alpha _{n})$, and there are no other
solutions.

To see this, notice that the condition $\tilde{v}_{+}=0$ leads to the weak
detonation as discussed above, while the condition $\tilde{v}_{-}=0$ leads to the weak deflagration. Hence, we must look for solutions with both $\tilde{v}
_{+}>0$ and $\tilde{v}_{-}>0$. The latter is possible if $\xi _{w}>c_{-}$,
so that we can use the rarefaction wave to fulfill the condition $\tilde{v}
=0 $ far behind the wall. The profile must also have a shock discontinuity
at some point $\xi _{\mathrm{sh}}>\xi _{w}$ in order to fulfill the boundary
condition $\tilde{v}=0$ far in front of the wall.
To construct the profile we have, on the one hand, the condition $\tilde{v
}_{-}\leq v_{\mathrm{rar}}( \xi _{w}) $, like in the weak detonation case. On the other hand, we have the condition $\tilde{v}
_{+}=\tilde{v}_{1} $ between $\xi _{w}$ and $\xi _{\mathrm{sh}}$, like in the weak deflagration case. For the bag EOS we have to look at  the left panel of
Fig. \ref{figsols} (case A). Then, we see that the above conditions give $\tilde{v}_{-}\leq v_{\mathrm{rar}}( \xi _{w}) <\tilde{v}_{1}
( \xi _{\mathrm{sh}}) =\tilde{v}_{+}$. This implies $v_{-}>v_{+}$, i.e., the solution must be a \emph{deflagration}. Besides, the condition $\tilde{v}_{-}\leq v_{\mathrm{rar}}( \xi _{w}) $ implies $v_{-}\geq c_{-}$. Since a strong
deflagration is unstable \cite{hkllm,mm14}, the only possibility here is $v_{-}=c_{-}$. Hence, this profile corresponds to a \emph{Jouguet deflagration}.

\begin{figure}[hbt]
\centering
\epsfysize=4cm \leavevmode \epsfbox{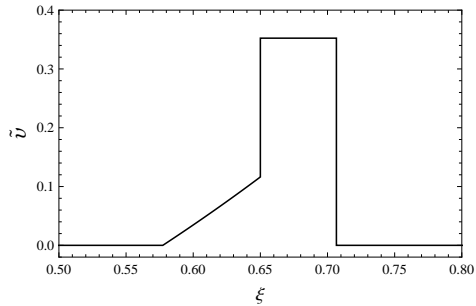}
\caption{The fluid velocity profile for a
supersonic Jouguet deflagration with velocity $v_w=0.65$, for the case $c_{+}=c_{-}=1/\sqrt{3}$, and for $\alpha _{n}=0.1$.}
\label{figjdefla}
\end{figure}

The profile is shown in Fig. \ref{figjdefla}. The Jouguet condition $v_{-}=c_{-}$ implies that $\tilde{v}_{-}= v_{\mathrm{rar}}( \xi _{w}) $, i.e., the rarefaction wave begins at the wall.  The value of $\tilde{v}_{-}$ is given by $\tilde{v}_{-}=(v_{w}-c_{-})/(1-v_{w}c_{-})$. The value
of $\tilde{v}_{+}$ is given by
\begin{equation}
\tilde{v}_{+}=\frac{v_{w}-v_{J}^{\mathrm{def}}(\alpha _{+})}{1-v_{w}v_{J}^{
\mathrm{def}}(\alpha _{+})},  \label{vmatildefj}
\end{equation}
and $\alpha _{+}$ is given by Eq. (\ref{a1an}) as a function of $\alpha _{n}$
and $\tilde{v}_{+}$. The two equations can be solved to obtain $\tilde{v}
_{+} $ as a function of $\alpha _{n}$ (see appendix \ref{flprof}). Then we can use
Eq. (\ref{xish}) to obtain the shock position $\xi _{\mathrm{sh}}$.
The enthalpy in the shock-wave region is
given, like in the weak deflagration case, by $w_{+}/w_{n}=\alpha _{n}/\alpha
_{+}$. On the other hand, the enthalpy behind the wall is given by Eq.
(\ref{enthprof}), with boundary conditions $\tilde{v}_{0}=\tilde{v}_{-}$ and $
w_{0}=w_{-}$ at $\xi =\xi _{w}$. The value of $w_{-}$ is related to $w_{+}$
through
\begin{equation}
w_{-}=\frac{v_{+}\gamma _{+}^{2}}{c_{-}\gamma ^{2}(c_{-})}w_{+}.
\label{wmewmaJ}
\end{equation}

As we increase the wall velocity, the wall position $\xi _{w}=v_{w}$
approaches the shock position $\xi _{\mathrm{sh}}$. As a consequence, the
maximum wall velocity for this kind of solution is obtained for $\xi
_{w}=\xi _{\mathrm{sh}}$. For case A, it can be shown that in this
limit the wall velocity coincides with that of the Jouguet detonation $
v_{J}^{\det }(\alpha _{n})$. Thus, supersonic Jouguet deflagrations
fill the gap between weak
deflagrations and weak detonations. This solution exists also
in cases B and C. However, in case C it does not fill the gap between $c_{-}$
and $v_{J}^{\det }(\alpha _{n})$, since there is still another kind of
solution.

\subsubsection{The subsonic detonation}

\begin{figure}[hbt]
\centering
\epsfysize=4cm \leavevmode \epsfbox{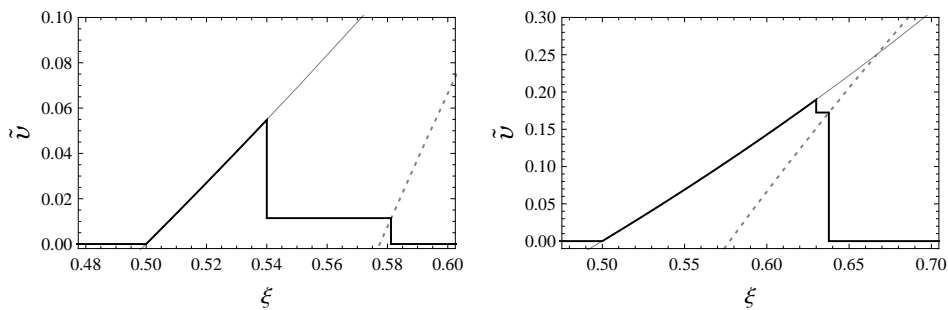}
\caption{The fluid velocity profiles for two subsonic detonations (solid lines) for the case $c_{+}=1/\sqrt{3}\simeq 0.58$, $c_{-}=0.5$. The grey lines indicate the functions $\tilde{v}_{\mathrm{rar}}(\xi )$
and $\tilde{v}_{1}(\xi _{\mathrm{sh}})$ corresponding to the right panel of
Fig. \ref{figsols}. The solution in the left panel corresponds to $\alpha _{n}=0.07$ and $v_{w}=0.54$, and the one in the right panel corresponds to $\alpha _{n}=0.1$ and $v_{w}=0.63$.}
\label{figjdetoc}
\end{figure}

As can be seen in the right panel of Fig. \ref{figsols}, in case C there is
a range of values of $\xi _{w}>c_{-}$ for which we may have $v_{\mathrm{rar}
}( \xi _{w}) >\tilde{v}_{1}( \xi _{\mathrm{sh}}) $,
provided that the shock velocity $\xi _{\mathrm{sh}}$ is close enough to $
c_{+}$ and the wall velocity $\xi _{w}$ is close enough to $c_{-}$. We may
thus construct a profile for which we have $\tilde{v}_{-}>\tilde{v}_{+}$, as
shown in Fig. \ref{figjdetoc} (we  show the lines of
Fig. \ref{figsols} for guidance). For such a profile we have $v_{-}<v_{+}$ and, hence, the solution is a \emph{detonation}. This detonation, however, is preceded by a shock wave. Physically, this is because, as we anticipated from Fig. \ref{figvmavmec}, in the case $c_{+}>c_{-}$ the incoming flow may be
subsonic ($v_{+}<c_{+}$) for parameters near the weak limit. In such a case,
the fluid in front of the wall is perturbed by the latter.
As can be observed in Fig. \ref{figjdetoc}, we have set $\xi _{0}=\xi _{w}$,
i.e., we have considered the Jouguet point, like in the previous case. Indeed, the condition $\tilde{v}_{-}\leq v_{\mathrm{rar}}( \xi _{w}) $
implies, as before, $v_{-}\geq c_{-}$, and it can be shown that the case $
v_{-}>c_{-}$ is  unstable\footnote{For $v_{-}>c_{-}$ and $v_{+}<c_{+}$, the conditions of the fluid on both sides of the wall are the same as in the
case of a strong deflagration. As a
consequence, the analysis of Ref. \cite{mm14}, which shows that such a front
is not evolutionary \cite{landau} applies to this case as well (in brief,
the total number of unstable modes is larger than the matching conditions at
the interface, and the solution is trivially unstable).}. Hence, the only
possibility is $v_{-}=c_{-}$, and we have a \emph{Jouguet detonation}.

This detonation moves with velocity $v_-=c_-$ with respect to the fluid just behind it, and with velocity $v_{J}^{\mathrm{\det }}(\alpha _{+})<c_+$ (i.e.,
subsonically) with respect to the fluid in front. Notice, however,
that the wall velocity $v_{w}$, like in the case of the supersonic
deflagration, is higher than both $v_{-}$ and $v_{+}$. Moreover, the
temperature variable $\alpha _{+}$ is not given by the value $\alpha _{n}$
anymore, since the fluid is reheated. As a consequence,
the wall is supersonic with respect to the bubble center, but
we may have either $v_{w}<c_{+}$, as in the left panel of Fig. \ref{figjdetoc}, or $v_{w}>c_{+}$ as in the right panel.

The calculation of the profile can be done like in the previous case. The
value of $\alpha _{+}$ is given by Eq. (\ref{a1an}), and the value of $
\tilde{v}_{+}$ is like in Eq. (\ref{vmatildefj}), replacing $v_{J}^{
\mathrm{def}}(\alpha _{+})$ with $v_{J}^{\mathrm{\det }}(\alpha _{+})$. The
rarefaction is obtained from the boundary condition $\tilde{v}
_{-}=(v_{w}-c_{-})/(1-v_{w}c_{-})$ at $\xi =\xi _{w}$. In appendix \ref{flprof}, the
profiles of the two Jouguet solutions (the supersonic deflagration and the
subsonic detonation) are obtained from a single calculation by writing $
\alpha _{+}$ as a function of $v_{+}$, which has the same expression for
detonations and deflagrations.

\begin{figure}[hbt]
\epsfysize=6cm \leavevmode \epsfbox{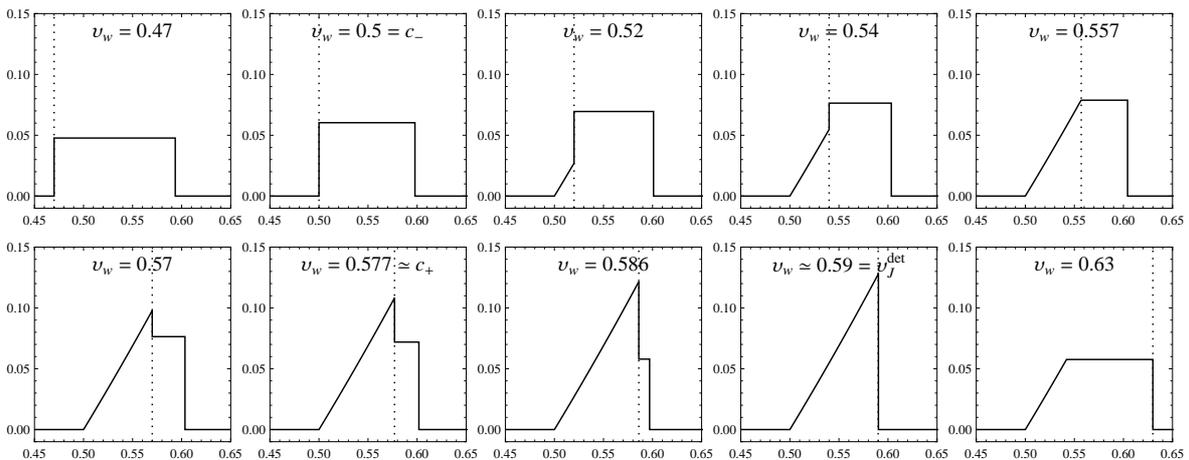}
\caption{Fluid velocity $\tilde v$ as a function of $\xi$ for the case $c_{+}=1/\protect\sqrt{3}$, $c_{-}=0.5$, for $\alpha_n=0.08$ and different wall velocities. The dotted lines indicate the phase interface.}
\label{figjouguet}
\end{figure}

In Fig. \ref{figjouguet} we plot several fluid velocity profiles for case C, for wall velocities in a range which includes the interval $[c_-,v_J^{\mathrm{det}}(\alpha_n)]$. We have considered a weak enough phase transition, i.e., $\alpha _{n}=0.08$, which is close to the weak limit (see Fig. \ref{figvmavmec}). Therefore, we have subsonic detonations, which do not exist for large values of $\alpha_n$. The first plot corresponds to a subsonic weak deflagration. The second solution corresponds to a wall moving at the speed of sound $c_-$, and is the limit between the weak and the Jouguet deflagrations. The third and forth profiles correspond to supersonic Jouguet deflagrations ($v_w>c_-$). These are subsonic, though, with respect to the speed of sound in the $+$ phase. The fifth case is the solution in the limit between the Jouguet deflagration and the Jouguet detonation. The subsequent three plots correspond to subsonic Jouguet detonations. These are supersonic with respect to the speed of sound $c_-$. The first of them is subsonic with respect to $c_+$, the second moves with  velocity $c_+$, and the last one is supersonic with respect to $c_+$. However, in the three cases the wall moves subsonically with respect to the fluid in front of it, $v_+=v_J^{\mathrm{det}}(\alpha_+)<c_+$. Hence, it is preceded by a shock front. The ninth solution is the limit between the Jouguet and weak detonations, and moves with velocity $v_J^{\mathrm{det}}(\alpha_n)$. Finally, the last plot corresponds to a weak detonation.

We see that Jouguet solutions are the strongest ones, i.e., those which
cause the largest disturbances in the fluid. Indeed, for weak deflagrations and detonations which are slower and faster, respectively, than the ones shown in Fig. \ref{figjouguet} we have smaller fluid velocities. The first profile shown in Fig. \ref{figjouguet} already corresponds to a relatively fast weak deflagration. As we increase the value of  $v_w=\xi_w$ and reach $\xi_w=c_-$, the rarefaction tail appears behind the wall. Thus, the profile of the weak deflagration transforms continuously into that of the supersonic Jouguet deflagration. As we increase the wall velocity further, the supersonic Jouguet deflagration  continuously transforms into the subsonic Jouguet detonation. Meanwhile, the values of $\xi_w$ and $\xi_\mathrm{sh}$ approach each other. Finally, as $v_w$ reaches the value $v_J^{\mathrm{det}}(\alpha_n)$, the phase transition front meets the shock front, and the subsonic Jouguet detonation transforms into the weak detonation. Although the profile seems to change continuously, the change is in fact discontinuous, since the velocity $\tilde v_+$ changes from a finite value to $\tilde v_+=0$. In the cases in which subsonic detonations do not exist, the supersonic Jouguet deflagrations transform into the weak detonation in a similar manner, as is well known from the case of the bag EOS.

\subsection{Efficiency factors}

We have argued that the efficiency factor $\kappa =E_{\mathrm{kin}}/\left(
\Delta \epsilon V_{b}\right) $ does not have a sensible physical
interpretation, and it may not even be useful in practice, since the quantity $\Delta
\epsilon $ is not easy to identify in general models.
The same happens to the natural variable of the bag EOS, $\alpha =\Delta \epsilon /(a_{+}T^{4})$, which in our model directly
generalizes to $\alpha =\Delta \epsilon /(a_{+}T^{\nu _{+}})$. The latter can be replaced by the more physical $L/w_{+}(T)$. Therefore, we will calculate
the factor  $\tilde{\kappa}=E_{v}/(\Delta e_{n}V_{b})$ as a function of physical parameters.
According to Eq. (\ref{kappatil}), we have
\begin{equation}
\tilde{\kappa}=\frac{w_{n}}{\Delta e_{n}}\frac{1}{v_{w}}I,  \label{ktil}
\end{equation}
where
\begin{equation}
I=\int_{0}^{\infty }\frac{w}{w_{n}}\tilde{v}^{2}\tilde{\gamma}^{2}d\xi .
\label{integ}
\end{equation}
We have normalized the enthalpy density to the boundary value $
w_{n}=w_{+}(T_{n})$ since it is proportional to this
value. Below we calculate the integral $I$, which depends on the fluid
profile. In this model, the released energy density $\Delta e_{n}=e_+(T_n)-e_-(T_n)$ can be expressed in terms of the amount of supercooling $T_{n}/T_{c}$ and the physical parameter $\bar{L}
=L/w_{+}(T_{c})$ as
\begin{equation}
\frac{\Delta e_{n}}{w_{n}}=\frac{1}{1+c_{+}^{2}}\left[\alpha _{c}\left(\frac{T_{c}}{T_{n}}\right)^{\nu _{+}}+1-\frac{
c_{+}^{2}-\alpha _{c}}{c_{-}^{2}}\left( \frac{T_{n}}{T_{c}}\right) ^{\nu
_{-}-\nu _{+}}\right] ,
\end{equation}
where the parameter $\alpha _{c}$ is given by
\begin{equation}
\alpha _{c}=c_{-}^{2}\frac{1+c_{+}^{2}}{1+c_{-}^{2}}\bar{L}+\frac{
c_{+}^{2}-c_{-}^{2}}{1+c_{-}^{2}}
\end{equation}
(for the bag EOS we have $\alpha _{c}=\bar{L}/3$), and $\nu_\pm=1+1/c_\pm^2$.

For weak deflagrations the integral in Eq. (\ref{ktil}) is trivial. We have
\begin{equation}
I=\frac{w_{+}}{w_{n}}\tilde{v}_{+}^{2}\tilde{\gamma}_{+}^{2}(v_{\mathrm{sh}
}-v_{w})\qquad \text{(weak deflagrations),}
\end{equation}
where we use the notation $\tilde{\gamma}_{\pm }^{2}=1/(1-\tilde{v}_{\pm
}^{2})$. The ratio $w_{+}/w_{n}$ is given by Eq. (\ref{a1an}).
The values of $\tilde{v}_{+}$ and $v_{\mathrm{sh}}$ are given in
appendix \ref{flprof} as functions of  $\alpha _{n}=\alpha_c(T_c/T_n)^{\nu_+}$ and $v_{w}$.

For weak detonations, the integrand in Eq. (\ref{integ}) is a
constant between $\xi _{0}$ and $\xi _{w}$, while in the  rarefaction region is obtained from  Eqs.
(\ref{vrar2}) and (\ref{enthprof}). Hence, the integral separates in
two parts. For this model and for the planar case, we obtain analytical
expressions even for the rarefaction part (see appendix \ref{intrar}). We have
\begin{eqnarray}
I &=&\frac{w_{-}}{w_{n}}\left[ \tilde{\gamma}_{-}^{2}\tilde{v}_{-}^{2}\left(
v_{w}-\xi _{0}\right) +\left( \frac{1-c_{-}}{1+c_{-}}\frac{1-\tilde{v}_{-}}{
1+\tilde{v}_{-}}\right) ^{\frac{c_{-}\nu _{-}}{2}}\frac{f\left( \xi
_{0}\right) -f\left( c_{-}\right) }{1+c_{-}^{2}}\right] \\
&&\qquad \qquad \qquad \qquad \qquad \qquad \qquad \text{(weak detonations)},
\nonumber
\end{eqnarray}
where the function $f(\xi )$ is given in appendix \ref{intrar} in terms of the hypergeometric function. 
The value of
$w_{-}$ is given by Eq. (\ref{wme}),
and the values of $v_{-}$, $\tilde{v}_{-}$, etc. are given in
appendix \ref{flprof}.

For the Jouguet solutions we have, similarly,
\begin{equation}
I=\frac{w_{+}}{w_{n}}\left[ \tilde{v}_{+}^{2}\tilde{\gamma}_{+}^{2}(v_{
\mathrm{sh}}-v_{w})+\frac{w_{-}}{w_{+}}\left( \frac{1-c_{-}}{1+c_{-}}\frac{1-
\tilde{v}_{-}}{1+\tilde{v}_{-}}\right) ^{\frac{c_{-}\nu _{-}}{2}}\frac{
f\left( v_{w}\right) -f\left( c_{-}\right) }{1+c_{-}^{2}}\right] ,
\end{equation}
with $w_{+}/w_{n}$ given by Eq. (\ref{a1an}) as a function of
$v_{\mathrm{sh}}$ and $w_{-}/w_{+}$ given by Eq. (\ref{wmewmaJ}).
The values of $v_{\mathrm{sh}}$, $v_{+}$, etc. can be expressed as functions
of $\tilde{v}_{+}$, which is given in appendix \ref{flprof} as a function of $\alpha_{n}$ and $v_w$.

\section{The efficiency factor and the speed of sound \label{effic}}

We have seen that, for a given set of thermodynamical parameters and a given
amount of supercooling, there is a hydrodynamical solution for any value of the wall velocity $v_w$.
However, in  a concrete problem the latter is not a free parameter.
The wall velocity depends
essentially on the difference of pressure between the two phases and on the
friction force of the wall with the plasma. In general, $v_w$ can be calculated as a function of the thermodynamical parameters, the temperature, and a friction parameter $\eta $. The dependence on these variables is not trivial, since the hydrodynamics causes an effective friction on the wall \cite{ms09,kn11,m01}. As a result, depending on the parameters, some values
of $v_{w}$ will never be realized, no matter the friction. Besides, for some sets of parameters there will be multiple hydrodynamic solutions, with different values of $v_w$. These issues have been extensively investigated for the bag EOS  (for a recent discussion, see \cite{ms12}). Since the hydrodynamics depends on the speed of sound, we expect that the wall velocity will depend on $c_\pm$ as well.
We shall address this issue elsewhere.
Below, we shall leave $v_w$ as a free parameter and investigate the disturbance of the plasma, ignoring the backreaction on the wall velocity.

In order to investigate the dependence of hydrodynamics on the value of
the speed of sound, we shall
compute the efficiency factor
$\tilde{\kappa}=E_{\mathrm{kin}}/(\Delta e_{n}V_{b})$, obtained in the previous section as a function of the parameters  $c_\pm$, $\bar{L}$, $T_{n}/T_{c}$ and $v_w$.

\begin{figure}[hbt]
\centering
\epsfysize=7cm \leavevmode \epsfbox{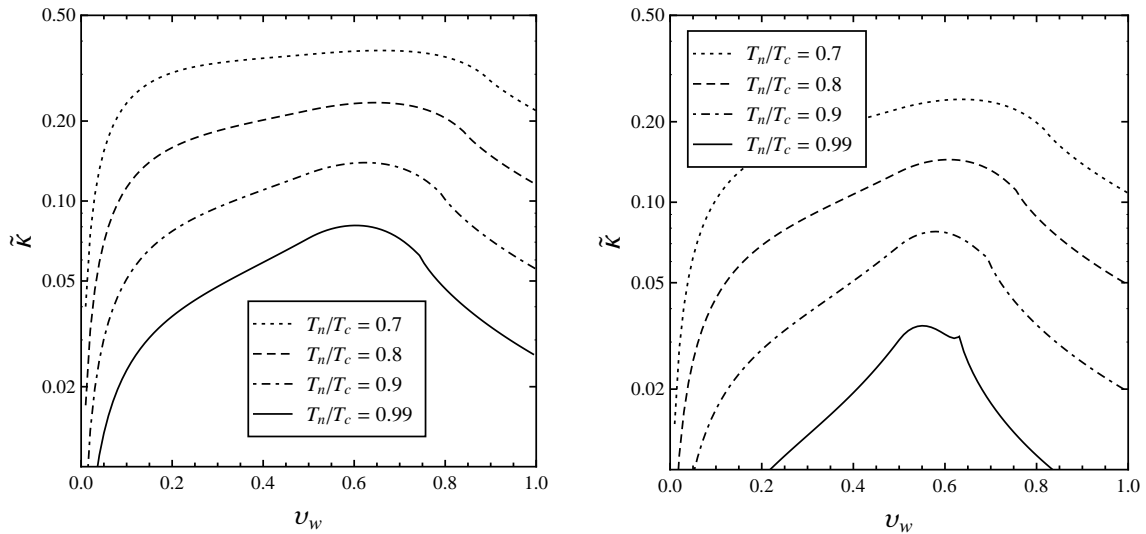}
\caption{The efficiency factor $\tilde{\kappa}$ as a function of the wall
velocity for the case $c_{+}=1/
\sqrt{3}$, $c_{-}=0.5$, and for $\bar{L}=0.5$ (left panel),
$\bar{L}=0.1$ (right panel).}
\label{figktilc}
\end{figure}

Let us begin by considering case C,
which is probably the most realistic case (at least, according to the
one-loop free energy, as discussed in Sec. \ref{EOSs}).
In Fig. \ref{figktilc} we show the result for two values of the latent heat parameter $\bar{L}$, for the case $c_{+}=1/\sqrt{3},c_{-}=0.5$. Qualitatively, the results are generally similar to the bag case\footnote{Notice that, for the bag case, the factor $\kappa$ considered elsewhere is related to our factor $\tilde{\kappa}$ by $\kappa \simeq \tilde 4\kappa$.} (see, e.g., Refs. \cite{ekns10,lm11}). Indeed, the
efficiency factor is small for weak deflagrations ($v_{w}<c_{-}$),
maximizes for Jouguet solutions ($c_{-}<v_{w}<v_{J}^{\det }$), and decreases again for weak detonations ($v_{w}>v_{J}^{\det }$). Besides, a larger latent heat or a stronger supercooling give higher values of the efficiency factor.

As can be appreciated in the right panel of Fig. \ref{figktilc},
for small latent heat and little supercooling there is a change in the behavior of $\tilde \kappa$ with $v_w$ near the maximum. A corner in the graphs appears and becomes more pronounced for $T_n/T_c$ closer to 1.
This corner is in fact present in all the curves (and in all the cases, including the bag case), only that it is less noticeable. This behavior
is due to the discontinuity between the profile of the weak detonation (for
which we have $\tilde{v}_+=0$, $T_{+}=T_{n}$) and that of the Jouguet solutions (for which we have $\tilde{v}_+>0$, $T_{+}>T_{n}$). The discontinuity in the kinetic energy density causes a corner in the graph of the efficiency factor\footnote{There will be also a jump in the dependence of the wall velocity on the parameters of the model \cite{lms12}.}. The corner becomes
more noticeable for weak phase transitions due to the appearance of the subsonic detonation. Relatively, this solution causes less disturbance of the fluid than the supersonic deflagration
(since we have $\tilde{v}_{+}<\tilde{v}_{-}$), and tends to lower the value
of $\tilde{\kappa}$.

\begin{figure}[hbt]
\centering
\epsfysize=7cm \leavevmode \epsfbox{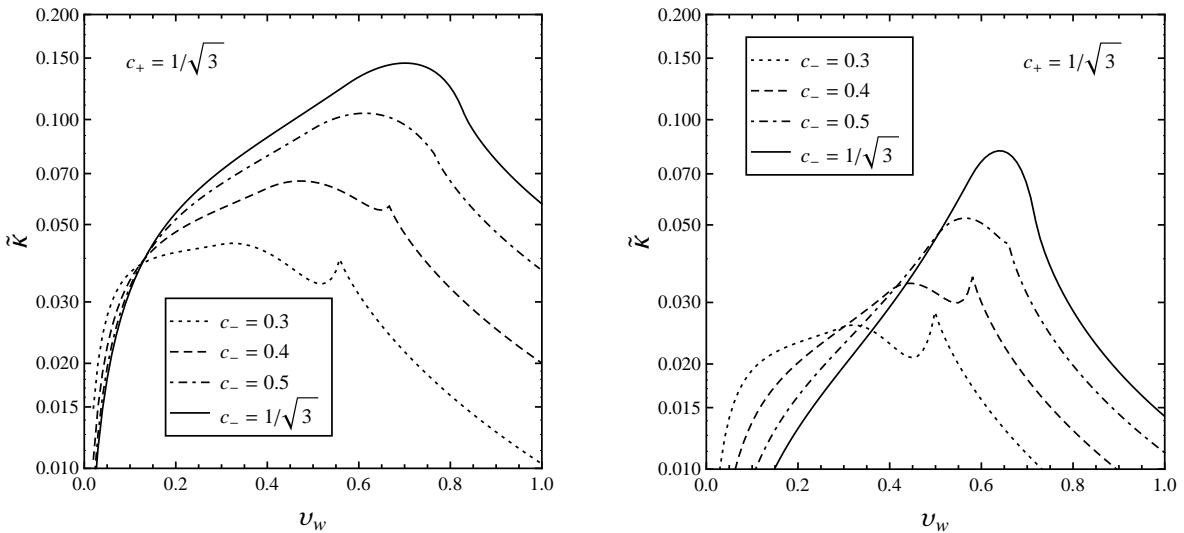}
\caption{The efficiency factor $\tilde{
\kappa}$ as a function of the wall velocity for the case $c_{+}=1/
\sqrt{3}$ and different values of $c_{-}$, for a supercooling of $T_{n}/T_{c}=0.95$
and for $\bar{L}=0.5$ (left panel) and $\bar{L}=0.1$ (right panel).}
\label{figktilc2}
\end{figure}

The qualitative and quantitative differences with the bag approximation increase as $c_-$ departs from $c_+$. In Fig. \ref{figktilc2} we consider a few cases with $c_{-}<c_+$, where the speed of sound in the high-temperature phase is that of radiation, $c_{+}=1/\sqrt{3}$.
The bag case ($c_{-}=c_{+}$) is plotted in a black solid line. We  observe that, for $c_{-}<1/\sqrt{3}$, the maximum efficiency is smaller than the bag result.
Besides, as $c_-$ decreases,  the peak of the curves moves to the left. This happens because the efficiency is always maximum between $c_{-}$
and $v_{J}^{\det}$, and this region moves to the left as $c_-$ decreases. As a consequence of this effect, the efficiency factor in the weak deflagration region is generally enlarged with respect to the bag case.
For small $c_-$ we see again the effect of the appearance of subsonic detonations, which lower the values of $\tilde{\kappa}$ in the
Jouguet region.

\begin{figure}[hbt]
\centering
\epsfysize=7cm \leavevmode \epsfbox{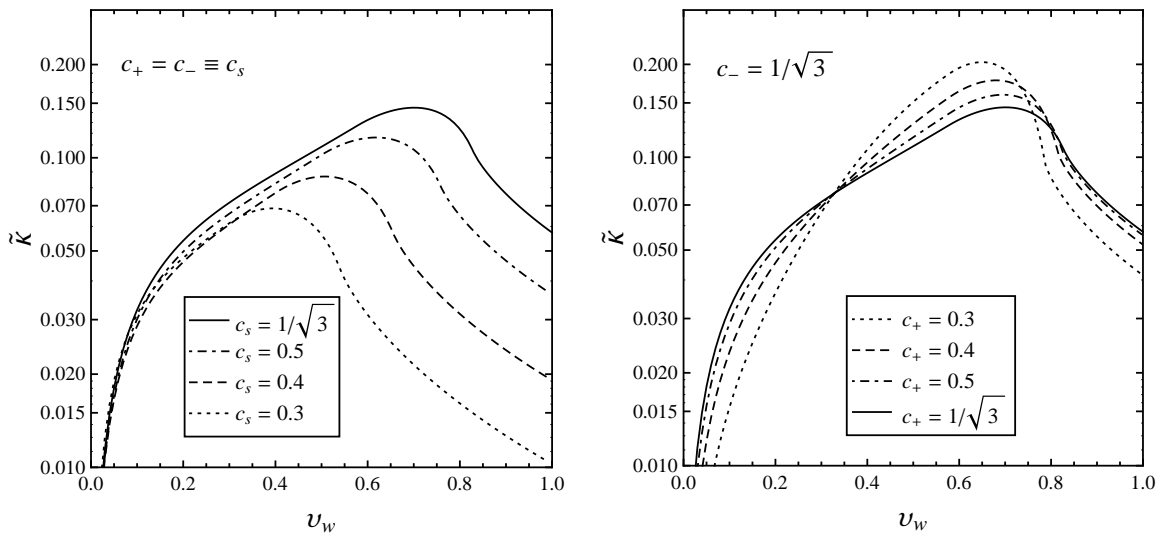}
\caption{The efficiency factor $\tilde{\kappa}$ as a function of the wall velocity, for $\bar{L}=0.5$, $T_{n}/T_{c}=0.95$, and several values of the speed of sound. The left panel corresponds to case A ($c_{+}=c_{-}$), and the right panel to case B ($c_{+}<c_{-}$).}
\label{figktilab}
\end{figure}

According to the discussion of Sec. \ref{EOSs}, cases A and B do not seem to be as likely as case C, but are certainly not impossible.
In the left panel of Fig. \ref{figktilab} we considered different values of the speed of sound for $c_{+}=c_{-}$ (case A). We observe that the position of the maximum gets displaced, as expected since it is always in the Jouguet region. Besides, we see that the production of kinetic energy is
more efficient for a fluid with a higher speed of sound.

In the right panel of Fig. \ref{figktilab} we considered some examples of case B. We fixed the speed of sound to the radiation value in the low-temperature phase and considered different values of $c_+$, with $c_+<c_-$. We observe that in this case the efficiency is larger than in the bag case for the range of wall velocities which maximizes the efficiency factor.
On the other hand, weak deflagrations or detonations generally give smaller efficiency factors than in the bag case. Nevertheless, we see that in this case the results do not depart significantly from those of the bag EOS, in contrast to what happens in the case $c_-<c_+$ (cf. Fig. \ref{figktilc2}).

In all these examples, we have limited the sound velocity in the two phases to values $c_s\leq 1/\sqrt{3}$. Considering values of $c_\pm$ higher than this bound gives generally higher values of the efficiency factor, and the curves of $\tilde \kappa$ vs. $v_w$ move to the right.

\section{Conclusions} \label{conclu}

Studying the propagation of phase transition fronts in a first-order phase
transition of a relativistic system is not an easy task. To simplify the
calculations, it is often necessary to make some approximations, such as
considering a simple equation of state.
A frequently used equation of state is the bag EOS, which often allows to obtain analytic or semi-analytic results. The bag EOS assumes that the two phases of the system consist only of radiation and vacuum energy.
In this work we have discussed, on the one hand, on the general capability of phenomenological equations of state such as the bag EOS to actually fit a given physical model. On the other hand, we have studied a specific characteristic of the plasma, namely, the speed of sound, which quantifies the departure from the bag EOS. Indeed, the latter gives the radiation value in both phases, $c_{-}=c_{+}=1/\sqrt{3}$.
Therefore, we have investigated the value of the speed of sound in physical models, as well as the dependence of hydrodynamics on this quantity.

In order to explore different physical models, we have considered
the one-loop finite-temperature effective potential for a system of particles which acquire masses through the Higgs mechanism. We have considered both analytical approximations and specific numerical examples (consisting of the electroweak phase transition for extensions of the Standard Model). We have seen that the speed of sound is bounded by the value $1/\sqrt{3}$, and we have  shown that in the
high-temperature phase the sound velocity $c_{+}$ is generally close to this value. However, in the low-temperature phase we may have values of the speed of sound as low as $c_{-}\lesssim 0.3$, depending on the model.

To study the dependence of the hydrodynamics on $c_\pm$, we have introduced a model which is the simplest generalization of the bag EOS and
incorporates the values of $c_{+}$ and $c_{-}$ as additional free parameters. As a consequence, our EOS includes the bag EOS as a particular case, and can give a better fit to a given physical model.
Thus, varying the parameters, the EOS can describe
different kinds of phase transitions, and the phase structure is more complex than in the bag case. For some values of the parameters, we may even have two phase transitions (i.e., two critical temperatures). However, as with the bag EOS, we do not expect that this simple EOS will describe a realistic model in a large temperature range, but only in a small range around a single critical temperature.

Our EOS preserves the computational simplicity of the bag EOS. Indeed, we
have seen that calculations with this model can be carried out much in the
same way as which is done in the bag case, and that analytic results can be
obtained as well.
We have obtained in particular the fluid profiles for the case of planar phase transition fronts, and we have calculated the fraction of the energy released in the phase transition which goes into bulk motions of the fluid.
As a result, we have found analytic equations for the efficiency factor $\tilde{\kappa}$ as a function of $c_\pm$, $v_w$, and the two parameter combinations $\alpha _{c}$ and $\alpha _{n}$. These latter can be expressed in terms of the more physical parameters $\bar{L} =L/w_{+}(T_{c})$ and $T_{n}/T_{c}$.

The fluid profiles have in general the same shape as in the bag case, with one exception. In the case $c_-<c_+$ and for low latent heat and little supercooling, a different solution may appear. The new solution is a Jouguet detonation which is subsonic with respect to the fluid in front of it. As a consequence, the bubble wall is preceded by a shock wave.
This solution is
supersonic with respect to the bubble center, and its fluid profile
is similar to that of the supersonic Jouguet deflagration. Moreover, as a function of $v_w$, the latter transforms continuously into the former.

Qualitatively, the  efficiency factor  $\tilde{\kappa}$ generally behaves like in the bag case as a function of the wall velocity and the thermodynamical parameters. However, some differences arise as $c_{+}$ and $c_{-}$ depart from the value $1/\sqrt{3}$.
In particular, for $c_-<c_+$ the maximum efficiency, which is obtained for Jouguet solutions, is smaller than in the bag case.
This means that the intensity of gravitational waves will be generally smaller than the results obtained using the bag EOS as an approximation. Quantitatively, these differences can be significant, depending on the parameters.

\section*{Acknowledgements}

This work was supported by Universidad Nacional de Mar del Plata,
Argentina, grant EXA699/14.

\appendix

\section{Calculation of fluid profiles} \label{flprof}

In this appendix we write down the equations for the parameters that enter
the fluid profiles, namely, $\tilde v_\pm$, $v_\mathrm{sh}$, $\xi_0$, and $\alpha_+$, as functions of $\alpha_n$ and $v_w$.

\paragraph{The weak detonation.}

The weak detonation profile is very easy to calculate, since there is no
shock wave reheating the plasma in front of the wall (i.e., $\tilde{v}_{+}=0$).
Therefore, we have $\alpha _{+}=\alpha _{n}$ and $v_{+}=v_{w}=\xi _{w}$.
We obtain $v_{-}$ from the same quadratic equation which gives Eq. (\ref{vmavme}). We have
\begin{equation}
v_{-}=\left( \frac{v_{+}\left( 1+\alpha _{+}\right) }{2q}+\frac{
c_{+}^{2}-\alpha _{+}}{2qv_{+}}\right) \pm \sqrt{\left( \frac{v_{+}\left(
1+\alpha _{+}\right) }{2q}+\frac{c_{+}^{2}-\alpha _{+}}{2qv_{+}}\right)
^{2}-c_{-}^{2}}.  \label{vmevma}
\end{equation}
As can be seen in Figs. \ref{figvmavmea},\ref{figvmavmeb},\ref{figvmavmec}, we have a gap around $v_+=c_-$, which indicates that the square root in Eq. (\ref{vmevma}) becomes imaginary. This gap separates the detonation branch from the deflagration branch. For each of these branches $v_-$ is a multivalued function of $v_+$. Thus,  the $\pm$ signs correspond to weak and strong solutions. At the Jouguet point the square root in Eq. (\ref{vmevma}) vanishes.
Weak detonations correspond to $v_+>c_-$ and to the $+$ sign.
In the frame of the bubble
center, the fluid velocity behind the wall is given by $\tilde{v}
_{-}=(v_{w}-v_{-})/(1-v_{w}v_{-})$. The rarefaction solution is matched at
the point $\xi _{0}$ given by $\xi _{0}=(\tilde{v}_{-}+c_{-})/(1+\tilde{v}
_{-}c_{-})$. The rarefaction ends at $\xi =c_{-}$. Behind this point, we
have $\tilde{v}=0$.

\paragraph{The weak deflagration.}

The weak deflagration solution has the simplest profile, namely, a constant $\tilde{v}_{+}=$ $\tilde{v}_{1}$ between the values $\xi =\xi _{w}$ and $\xi
=\xi _{\mathrm{sh}}$, and $\tilde v=0$ elsewhere. However, the calculation of these parameters is more involved than the detonation case, since we have to consider the matching conditions at the wall as well as at the shock front. In this case we have $v_-=v_w$, and we may relate $\tilde{v}_{+}$ to $v_w$ and $\alpha_+$ using Eq. (\ref{vmavme}). On the other hand, we may relate $\tilde{v}_{1}$ to $\alpha _{+}$ and $\alpha _{n}$ using Eq. (\ref{a1an}). From $\tilde{v}_{+}=$ $\tilde{v}_{1}$ we may eliminate $\alpha _{+}$. We find it easier to use instead
the following expression for $\alpha _{+}$ as a function of $v_{+}$ and $
v_{-}$,
\begin{equation}
\alpha _{+}=\gamma _{+}^{2}\left[ v_{+}^{2}+c_{+}^{2}-q\left(
v_{+}v_{-}+c_{-}^{2}\frac{v_{+}}{v_{-}}\right) \right] ,  \label{amas2}
\end{equation}
which comes from the conditions (\ref{land1}-\ref{land2}) and is
equivalent to Eq. (\ref{vmavme}) and to Eq. (\ref{vmevma}).
The condition $\alpha_+=\alpha_1$ gives another expression for $\alpha_+$, namely, Eq. (\ref{a1an}),
$\alpha _{+}=\alpha _{n}{c_{+}^{2}(1-v_{\mathrm{sh}}^{2})}/({v_{\mathrm{sh}
}^{2}-c_{+}^{4}})$. Therefore, we can readily eliminate $\alpha_+$ in Eq. (\ref{amas2}). Writing $v_+$ as a function of $\tilde v_+$ from  Eq. (\ref{vmametil}), and $v_{\mathrm{sh}}$ as a function of $\tilde v_1$ from Eq. (\ref{xish}),
we obtain $\alpha_n$ as a function of $v_w$ and $\tilde v_+$,
\begin{equation}
\alpha _{n}=\frac{q\left[ \frac{1-c_{-}^{2}c_{+}^{2}}{1+c_{+}^{2}}v_{w}
\tilde{v}_{+}^{2}+\tilde{v}_{+}\left( c_{-}^{2}-v_{w}^{2}\right) +\frac{
c_{+}^{2}-c_{-}^{2}}{1+c_{+}^{2}}v_{w}\right] }{v_{w}\left[ 1+\frac{
1+c_{+}^{4}}{2c_{+}^{2}}\tilde{v}_{+}^{2}-\frac{1+c_{+}^{2}}{c_{+}}\tilde{v}
_{+}\sqrt{1+\frac{(1-c_{+}^{2})^{2}}{4c_{+}^{2}}\tilde{v}_{+}^{2}}\right] },
\end{equation}
\linebreak which can be solved to obtain $\tilde{v}_{+}$ as a function of $
\alpha _{n}$ and $v_{w}$. The value of $\xi _{\mathrm{sh}}$ is then obtained
from Eq. (\ref{xish}).

\paragraph{The Jouguet solutions.}

The shape of the supersonic deflagration and the subsonic detonation are
similar, the only difference being that the velocity $v_{+}$ is given by $
v_{J}^{\mathrm{def}}(\alpha _{+})$ and $v_{J}^{\det }(\alpha _{+})$,
respectively. Nevertheless, Eq. (\ref{amas2})  allows to treat the two Jouguet solutions at the same time.
Indeed, notice that this expression is valid either for deflagrations or
detonations. It is by inverting this equation that two solutions appear
for $v_{+}$ as a function of $v_{-}$ and $\alpha _{+}$.
In the Jouguet case, Eq. (\ref{amas2}) gives
\begin{equation}
\alpha _{+}=\gamma _{+}^{2}\left( v_{J}^{2}+c_{+}^{2}-2qc_{-}v_{J}\right) ,
\end{equation}
which is equivalent to both Eqs. (\ref{vj}) (i.e., by inverting this expression
we obtain $v_{J}^{\QATOP{\mathrm{det}}{\mathrm{def}}}$ as functions of $
\alpha _{+}$). Proceeding as before, we obtain
\begin{equation}
\alpha _{n}=\frac{\left[ (\tilde{v}_{+}-v_{w})^{2}+c_{+}^{2}(1-\tilde{v}
_{+}v_{w})^{2}+2qc_{-}(\tilde{v}_{+}-v_{w})(1-\tilde{v}_{+}v_{w})\right] }{
(1-v_{w}^{2})\left[ 1+\frac{1+c_{+}^{4}}{2c_{+}^{2}}\tilde{v}_{+}^{2}-\frac{
1+c_{+}^{2}}{c_{+}}\tilde{v}_{+}\sqrt{1+\frac{(1-c_{+}^{2})^{2}}{4c_{+}^{2}}
\tilde{v}_{+}^{2}}\right] },
\end{equation}
which gives $\tilde{v}_{+}$ as a function of $\alpha _{n}$ and $v_{w}$. The
shock position is then obtained from Eq. (\ref{xish}), taking into account
that $\tilde{v}_{1}=\tilde{v}_{+}$. The rarefaction wave begins at $\xi
_{0}=\xi _{w}$, with the value $\tilde{v}
_{-}=(v_{w}-c_{-})/(1-v_{w}c_{-})$.

\section{Kinetic energy integral for the rarefaction wave} \label{intrar}

In this appendix we find the integral of the (normalized) kinetic energy density, Eq. (\ref{integ}),
in the rarefaction region,
\begin{equation}
I_{\mathrm{rar}}\equiv \int_{c_{-}}^{\xi _{0}}\frac{w_{\mathrm{rar}}}{w_{n}}
\frac{\tilde{v}_{\mathrm{rar}}^{2}}{1-\tilde{v}_{\mathrm{rar}}^{2}}.
\end{equation}
According to Eqs. (\ref{vrar2}) and (\ref{enthprof}), we have
\begin{equation}
I_{\mathrm{rar}}=\frac{w_{-}}{w_{n}}\left[ \frac{1}{1-c_{-}^{2}}\left( \frac{
1-c_{-}}{1+c_{-}}\frac{1-\tilde{v}_{-}}{1+\tilde{v}_{-}}\right) ^{\frac{
c_{-}\nu _{-}}{2}}\bar{I}\right] ,
\end{equation}
where
\begin{equation}
\bar{I}=\int_{c_{-}}^{\xi _{0}}\left( \frac{1+\xi }{1-\xi }\right) ^{\frac{
c_{-}\nu _{-}}{2}}\frac{\left( \xi -c_{-}\right) ^{2}}{1-\xi ^{2}}d\xi .
\end{equation}
After the change of variable $x=\left( 1+\xi \right) /\left( 1-\xi \right) $,
the integral $\bar{I}$ becomes
\begin{equation}
\bar{I}=\int_{x_-}^{x_0} x^{\mu -1}\left( \frac{(1-c_{-})^{2}}{2}-\frac{2}{x+1}+\frac{2}{
(x+1)^{2}}\right) dx,
\end{equation}
with $\mu =c_{-}\nu _{-}$. Thus, the integral $\bar I$ splits into three integrals. The first one is trivial, and the other two can be expressed in terms of the hypergeometric functions\footnote{See \cite{grads}, Eq. 3.194-1.} $_{2}F_{1}(1,\mu ;\mu +1;-x)$ and $_{2}F_{1}(2,\mu
;\mu +1;-x)$, where $_{2}F_{1}\equiv F$ is defined as \cite{grads}
\begin{equation}
F(\alpha ,\beta ;\gamma ;z)=1+\frac{\alpha \beta }{\gamma 1}z+\frac{\alpha
(\alpha +1)\beta (\beta +1)}{\gamma (\gamma +1)1\cdot 2}z^{2}+\cdots .
\end{equation}
The hypergeometric functions with  $\alpha =1$ and $\alpha=2$ are in fact
related\footnote{Using Eq. 9.137-2 of Ref. \cite{grads} and the fact that
$F(0,\beta ;\gamma
;z)=1$.}, and we also have a relation\footnote{See \cite{grads}, Eq. 9.131-1.}
between $F(1,\mu ;\mu +1;-x)$ and $F(1,1;\mu
+1;\frac{x}{x+1})$. We thus obtain
\begin{equation}
\bar{I}=\frac{1-c_{-}^{2}}{1+c_{-}^{2}}\left[ f\left( \xi _{0}\right)
-f\left( c_{-}\right) \right] ,
\end{equation}
where
\begin{equation}
f\left( \xi \right) =\left[ \frac{1+\xi }{1-\xi }\right] ^{\frac{c_{-}\nu
_{-}}{2}}\left[ 1-\frac{1+c_{-}^{2}}{1-c_{-}^{2}}(\xi -c_{-})-\left( 1-\xi
\right) F\left(1,1;\frac{(c_{-}+1)^{2}}{2c_{-}};\frac{1+\xi }{2}\right)\right] .
\end{equation}
As a result, we have
\begin{equation}
I_{\mathrm{rar}}=\frac{w_{-}}{w_{n}}\left[ \left( \frac{1-c_{-}}{1+c_{-}}
\frac{1-\tilde{v}_{-}}{1+\tilde{v}_{-}}\right) ^{\frac{c_{-}\nu _{-}}{2}}
\frac{1}{1+c_{-}^{2}}\left[ f\left( \xi _{0}\right) -f\left( c_{-}\right)
\right] \right] .
\end{equation}

\end{document}